

\font\titlefont = cmr10 scaled\magstep 4
\font\sectionfont = cmr10
\font\littlefont = cmr5
\font\eightrm = cmr8 

\def\ss{\scriptstyle} 
\def\sss{\scriptscriptstyle} 

\magnification = 1200

\global\baselineskip = 1.2\baselineskip
\global\parskip = 4pt plus 0.3pt
\global\abovedisplayskip = 18pt plus3pt minus9pt
\global\belowdisplayskip = 18pt plus3pt minus9pt
\global\abovedisplayshortskip = 6pt plus3pt
\global\belowdisplayshortskip = 6pt plus3pt


\def\endignore{}
\def\ignore #1\endignore{}

\newcount\dflag
\dflag = 0


\def\monthname{\ifcase\month
\or Jan \or Feb \or Mar \or Apr \or May \or June%
\or July \or Aug \or Sept \or Oct \or Nov \or Dec
\fi}

\def\timestring{{\count0 = \time%
\divide\count0 by 60%
\count2 = \count0
\count4 = \time%
\multiply\count0 by 60%
\advance\count4 by -\count0
\ifnum\count4 < 10 \toks1 = {0}
\else \toks1 = {} \fi%
\ifnum\count2 < 12 \toks0 = {a.m.}
\else \toks0 = {p.m.}
\advance\count2 by -12%
\fi%
\ifnum\count2 = 0 \count2 = 12 \fi
\number\count2 : \the\toks1 \number\count4%
\thinspace \the\toks0}}

\def\today{\ifcase\month\or January\or February\or March\or
 April\or May\or June\or July\or August\or September\or
 October\or November\or December\fi \space\number\day, \number\year}



\def\endtitle{}
\def\title#1\endtitle{\vskip.5in\titlefont
\global\baselineskip = 2\baselineskip
#1\vskip.4in
\baselineskip = 0.5\baselineskip\rm}
 
\def\endauthors{}
\def\authors#1\endauthors{#1}

\def\endabstract{}
\def\abstract#1\endabstract{\vskip .3in%
\centerline{\sectionfont\bf Abstract}%
\vskip .1in
\noindent#1}

\newcount\nsection
\newcount\nsubsection

\def\section#1{\global\advance\nsection by 1
\nsubsection=0
\bigskip\noindent\centerline{\sectionfont \bf \number\nsection.\ #1}
\bigskip\rm\nobreak}

\def\subsection#1{\global\advance\nsubsection by 1
\bigskip\noindent\sectionfont \sl \number\nsection.\number\nsubsection)\
#1\bigskip\rm\nobreak}

\def\topic#1{{\medskip\noindent $\bullet$ \it #1:}} 
\def\endtopic{\medskip}

\def\appendix#1#2{\bigskip\noindent%
\centerline{\sectionfont \bf Appendix #1.\ #2}
\bigskip\rm\nobreak}


\newcount\nref
\global\nref = 1

\def\ref#1#2{\xdef #1{[\number\nref]}
\ifnum\nref = 1\global\xdef\therefs{\noindent[\number\nref] #2\ }
\else
\global\xdef\oldrefs{\therefs}
\global\xdef\therefs{\oldrefs\vskip.1in\noindent[\number\nref] #2\ }%
\fi%
\global\advance\nref by 1
}

\def\listrefs{\vfill\eject\section{References}\therefs}


\newcount\nfoot
\global\nfoot = 1

\def\foot#1#2{\xdef #1{(\number\nfoot)}
\footnote{${}^{\number\nfoot}$}{\eightrm #2}
\global\advance\nfoot by 1
}


\newcount\nfig
\global\nfig = 1

\def\fig#1{\xdef #1{(\number\nfig)}
\global\advance\nfig by 1
}


\newcount\cflag
\newcount\nequation
\global\nequation = 1
\def\eqlabel{(1)}

\def\nexteqno{\ifnum\cflag = 0
\global\advance\nequation by 1
\fi
\global\cflag = 0
\xdef\eqlabel{(\number\nequation)}}

\def\lasteqno{\global\advance\nequation by -1
\xdef\eqlabel{(\number\nequation)}}

\def\label#1{\xdef #1{(\number\nequation)}
\ifnum\dflag = 1
{\escapechar = -1
\xdef\draftname{\littlefont\string#1}}
\fi}

\def\clabel#1#2{\xdef\eqlabel{(\number\nequation #2)}
\global\cflag = 1
\xdef #1{\eqlabel}
\ifnum\dflag = 1
{\escapechar = -1
\xdef\draftname{\string#1}}
\fi}

\def\cclabel#1#2{\xdef\eqlabel{#2)}
\global\cflag = 1
\xdef #1{\eqlabel}
\ifnum\dflag = 1
{\escapechar = -1
\xdef\draftname{\string#1}}
\fi}


\def\eeq{}

\def\eqnn #1\eeq{$$ #1 $$}

\def\eq #1\eeq{\xdef\draftname{\ }
$$ #1
\eqno{\eqlabel \rlap{\ \draftname}} $$
\nexteqno}



\def\eol{& \eqlabel \rlap{\ \draftname} \crcr
\nexteqno
\xdef\draftname{\ }}

\def\eeol{& \eqlabel \rlap{\ \draftname}
\nexteqno
\xdef\draftname{\ }}

\def\eolnn{\cr
\global\cflag = 0
\xdef\draftname{\ }}

\def\eeolnn{\xdef\draftname{\ }}

\def\eqa #1\eeq{\xdef\draftname{\ }
$$ \eqalignno{ #1 } $$
\global\cflag = 0}


\def\ie{{\it i.e.\/}}
\def\eg{{\it e.g.\/}}
\def\etc{{\it etc.\/}}
\def\etal{{\it et.al.\/}}


\def\plb#1#2#3{{\it Phys.~Lett.} {\bf #1B} (19#2) #3}

\def\prd#1#2#3{{\it Phys.~Rev.} {\bf D#1} (19#2) #3}
\def\pr#1#2#3{{\it Phys.~Rev.} {\bf #1} (19#2) #3}

\def\prl#1#2#3{{\it Phys.~Rev.~Lett.} {\bf #1} (19#2) #3}


\global\nulldelimiterspace = 0pt



\def\frac#1#2{{{#1} \over {#2}}\,}  
\def\hf{{1\over 2}}
\def\nth#1{{1\over #1}}


\def\Dsl{\hbox{/\kern-.6700em\it D}} 
\def\dsl{\hbox{/\kern-.5300em$\partial$}}
\def\pxpsl{\hbox{/\kern-.5600em$p$}}
\def\ssl{\hbox{/\kern-.5300em$s$}}
\def\epssl{\hbox{/\kern-.5100em$\epsilon$}}
\def\delsl{\hbox{/\kern-.6300em$\nabla$}}
\def\lxpsl{\hbox{/\kern-.4300em$l$}}
\def\elxpsl{\hbox{/\kern-.4500em$\ell$}}
\def\kxpsl{\hbox{/\kern-.5100em$k$}}
\def\qxpsl{\hbox{/\kern-.5000em$q$}}
\def\transp#1{#1^{\sss T}} 
\def\sla#1{\raise.15ex\hbox{$/$}\kern-.57em #1}
\def\Pl{\gamma_{\sss L}}
\def\Pr{\gamma_{\sss R}}
\def\pwr#1{\cdot 10^{#1}}



\def\roughly#1{\mathrel{\raise.3ex\hbox{$#1$\kern-.75em\lower1ex\hbox{$\sim$}}}}
\def\lsim{\roughly<}
\def\gsim{\roughly>}

\def\tw#1{\tilde{#1}}
\def\ol#1{\overline{#1}}



\def\bfp{{\bf p}}

\def\bfr{{\bf r}}

\def\bfx{{\bf x}}


\def\Bfd{{\bf D}}

\def\Bfj{{\bf J}}

\def\Bfp{{\bf P}}

\def\Bfr{{\bf R}}


\def\Sca{{\cal A}}

\def\Scl{{\cal L}}
\def\Scm{{\cal M}}

\def\Scq{{\cal Q}}

\def\Scv{{\cal V}}



\def\bra#1{\langle #1 |}
\def\ket#1{| #1 \rangle}

\def\vev#1{\langle 0 | #1 | 0 \rangle}

\def\Avg#1{\left\langle #1 \right\rangle}



\def\cc{{\rm c.c.}}

\def\sm{standard model}

\def\eV{{\rm \ eV}}
\def\keV{{\rm \ keV}}
\def\MeV{{\rm \ MeV}}


\def\sss{\scriptscriptstyle} 
\def\ss{\scriptstyle} 
\def\ngb{Nambu-Goldstone boson}
\def\gwk{$SU_{\sss L}(2) \times U_{\sss Y}(1)$}
\def\gfl{$SU_{\sss F}(2) \times U_{\sss L'}(1)$}
\def\ul{$U_{\sss L}(1)$}
\def\transp{{\sss T}}
\def\G{\Gamma}
\def\pf{p_{\sss F}}
\def\EF{E_{\sss F}}
\def\wf{w_{\sss F}}
\def\wgt{w_{\sss GT}}
\def\GF{G_{\sss F}}

\def\bb{\beta\beta} 
\def\bbtn{\bb_{2\nu}} 
\def\bbm{\bb_{m}}
\def\bbom{\bb_{om}} 
\def\bbcm{\bb_{cm}} 
\def\bbzn{\bb_{0\nu}}
\def\sm{s_{\sss -}}
\def\sp{s_{\sss +}}
\def\gm{g_{\sss -}}

\def\gp{g_{\sss +}}
\def\Mm{M_{\sss -}}
\def\Mp{M_{\sss +}}

\def\etal{{\it et al.}}

\def\Pr{\gamma_{\sss R}}
\def\Pl{\gamma_{\sss L}}

\def\vev#1{\left\langle #1\right\rangle}

\def\gtwid{\mathrel{\raise.3ex\hbox{$>$\kern-.75em\lower1ex\hbox{$\sim$}}}}
\def\ltwid{\mathrel{\raise.3ex\hbox{$<$\kern-.75em\lower1ex\hbox{$\sim$}}}}

\def\frac#1#2{{\scriptstyle{#1 \over #2}}}
\def\slp{{\raise.15ex\hbox{$/$}\kern-.57em\hbox{$\partial$}}}
\def\darr{\raise1.5ex\hbox{$\leftrightarrow$}\mkern-16.5mu \slp}
\def\gv{g_{\sss V}}
\def\ga{g_{\sss A}}

\def\bfp{{\bf p}}
\def\bfr{{\bf r}}
\def\bfrnm{{\bf r}_{\!nm}}

\def\ellsl{{\raise.15ex\hbox{$/$}\kern-.57em\hbox{$\ell$}}}
\def\psl{{\raise.15ex\hbox{$/$}\kern-.57em\hbox{$p$}}}
\def\qsl{{\raise.15ex\hbox{$/$}\kern-.57em\hbox{$q$}}}
\def\wtM{{\widetilde M}}
\def\Nd{$^{150}$Nd}
\def\Mo{$^{100}$Mo}
\def\Se{$^{82}$Se}
\def\Nd{$^{150}$Nd}
\def\U{$^{238}$U}
\def\Ge{$^{76}$Ge}
\def\Te#1{$^{#1}$Te}

\def\sigm{{\bf\sigma}_m}
\def\sn{{\bf\sigma}_n}

\def\spa{\!\phantom{-1}}
\def\SN#1#2{#1\times 10^{#2}}
\def\SNt#1#2{$#1\times 10^{#2}$}
\def\pr#1{Phys.~Rev.~{\bf #1}}
\def\np#1{Nucl.~Phys.~{\bf #1}}
\def\pl#1{Phys.~Lett.~{\bf #1}}
\def\prl#1{Phys.~Rev.~Lett.~{\bf #1}}

\def\plb#1#2#3{{ Phys.~Lett.~}{\bf #1B} (19#2) #3}

\def\prd#1#2#3{{ Phys.~Rev.~}{\bf D#1} (19#2) #3}
\def\prl#1#2#3{{ Phys.~Rev.~Lett.~}{\bf #1} (19#2) #3}
\def\ie{{\it i.e.}}
\def\eg{{\it e.g.}}
\def\etc{{\it etc.}}
\def\ol#1{\overline{#1}}
\def\geff{g_{\rm eff}}

\def\roughlyup#1{\mathrel{\raise.3ex\hbox{$\sim$\kern-.75em
\lower1ex\hbox{$#1$}}}}
\def\roughlydown#1{\mathrel{\raise.3ex\hbox{$#1$\kern-.75em
\lower1ex\hbox{$\sim$}}}}
\def\bra{\langle}
\def\ket{\rangle}
\def\lsim{\roughlydown<}
\def\gsim{\roughlydown>}
\def\simeq{\roughlyup-}

\def\dket{\rangle\!\rangle}
\def\dbra{\langle\!\langle}
\def\wtM{\widetilde M}


\rightline{McGill/93-02}
\rightline{TPI-MINN-93/28-T}
\rightline{UMN-TH-1204-93}
\rightline{June 1993}
\vskip .1in

\title
\centerline{A New Class of Majoron-Emitting}
\centerline{Double-Beta Decays}
\endtitle

\authors
\centerline{C.P. Burgess${}^a$ and J.M. Cline${}^b$\footnote{}{email:
cliff@physics.mcgill.ca; jcline@mnhep.hep.umn.edu}}  
\vskip .15in 
\centerline{\it ${}^a$ Physics Department, McGill University}
\centerline{\it 3600 University St., Montr\'eal, Qu\'ebec, CANADA, H3A 2T8.}
\vskip .1in 
\centerline{\it ${}^b$ Theoretical Physics Institute, 
The University of Minnesota}
\centerline{\it Minneapolis, MN, 55455, USA.}
\endauthors

\abstract
Motivated by the excess events that have recently been found near the endpoints
of the double beta decay spectra of several elements, we re-examine models in
which double beta decay can proceed through the neutrinoless emission of
massless Nambu-Goldstone bosons (majorons).   Noting that models proposed to
date for this process must fine-tune either a scalar mass or a VEV to be
less than 10 keV, we introduce a new kind of majoron which avoids this
difficulty by carrying lepton number $L=-2$. We analyze in detail the
requirements that models of both the conventional and our new type must satisfy
if they are to account for the observed excess events. We find: (1) the electron
sum-energy spectrum can be used to distinguish the two classes of models from
one another; (2) the decay rate for the new models depends on different nuclear
matrix elements than for ordinary majorons; and (3) all models require a
(pseudo) Dirac neutrino, having a mass of a several hundred MeV, which mixes
with $\nu_e$. 
\endabstract 


\vfill\eject
\section{Introduction and Summary}  

Recently, a mysterious excess of high-energy electrons has been seen in the
electron spectrum for the double-beta ($\bb$) decay of several elements.   This
kind of observation was first made in 1987 for the decay $^{76}$Ge $\to^{76}$Se
$+2e^-$ by Avignone \etal\ 
\ref\Avignone{F.T.~Avignone III \etal, in {\it Neutrino Masses and Neutrino
Astrophysics,}  proceedings of the IV Telemark Conference, Ashland, Wisconsin,
1987,  edited by V.~Barger, F.~Halzen, M.~Marshak and K.~Olive   (World
Scientific, Sinagpore, 1987), p.~248.}
\Avignone, although the effect was discounted when they, as well as other
groups,  subsequently excluded a signal having the original strength 
\ref\nomajoron{P.~Fisher \etal, \pl{B192} (1987) 460; D.O.~Caldwell \etal,
\prl{59}{87}{1649}.}
\nomajoron. The mysterious events are back, however, with the UC Irvine group
now finding excess numbers of electrons near but below the endpoints for \Mo,
\Se\ and \Nd, with a statistical  significance of 5$\sigma$ 
\ref\Moe{M.~Moe, M. Nelson, M. Vient and S.  Elliott, preprint UCI-NEUTRINO
92-1; {\it Nucl. Phys.} (Proc. Suppl.) {\bf B31} (1993).}
\Moe. Such events also persist in the ${}^{76}$Ge data 
\ref\ge{F.T.~Avignone III \etal, \pl{B256} (1991) 559.}
\ref\priv{F.T. Avignone, private communication.}
\ge, \priv, at approximately a tenth of the  original rate.  

Since these are difficult experiments, it is possible that  the anomalous
events will turn out to be due to systematic error, or to a hitherto
unsuspected nuclear physics effect. But they may also be the fingerprint of a
new fundamental interaction 
\ref\CMP{Y.~Chikashige,  R.N.~Mohapatra and R.D.~Peccei, \prl{45}{80}{1926};
\pl{98B} (1981) 265.}
\ref\GR{G.B.~Gelmini and M.~Roncadelli, \pl{99B} (1981) 411.}
\CMP, \GR, in which two nucleons decay with the emission of two electrons
and a light scalar, the majoron,\foot\ftone{The term ``majoron'' was
originally used for the \ngb\ associated with spontaneous  breaking of lepton
number, since the same lepton number breaking induced a Majorana mass for the
neutrinos. We enlarge the meaning of the name in this paper by applying it 
even if the scalar is massive or if the model in question does not generate
Majorana masses.}\  rather than the usual two-electron, two-neutrino
decay
\ref\Doi{M.~Doi, T.~Kotani and E.~Takasugi, \pr{D37} (1988) 2572.}
\ref\georgi{H.M.~Georgi, S.L.~Glashow  and S.~Nussinov, \np{B193} (1981)
297.}
\Doi, \georgi. If so, these observations are of vital importance since they
provide us with a glimpse of physics beyond the standard electroweak theory.   

We assume for the sake of argument that the excess high-energy electrons are
really due to majoron emission, denoted by $\bbm$. Our goal is to explore the
implications of $\bbm$ taken together with the other known constraints on
neutrino physics.  In so doing, we have found that the candidate models capable
of describing majoron emission from nuclei fall into two broad classes.  

In the first class of models for $\bbm$ --- which to our knowledge includes
everything that has been proposed until recently 
\ref\berezhiani{Z.G.~Berezhiani, A.Yu.~Smirnov and J.W.F. Valle,
\plb{291}{92}{99}.} 
\GR, \berezhiani\ --- the majoron is the \ngb\ associated with the spontaneous
breaking of a $U(1)$ lepton number symmetry. The only way to get an observable
rate in this context is to have either a scalar mass or vacuum expectation value
({VEV}) of the order of $10$ keV.  We refer to these as `ordinary' majoron
models (OMM's), and denote their associated beta decay by $\bbom$. We provide
here a first comprehensive analysis of which OMM's can give a large enough rate
of $\bbom$.  

In addition, we have recently proposed
\ref\ourletter{C.P.~Burgess and J.M.~Cline, Phys.~Lett.~{\bf B298} (1993)
141.}
\ourletter\ a second, qualitatively different, sort of majoron that does not
require such a small scale. Unlike OM's, this new majoron carries a classically
unbroken lepton number charge, and is the \ngb\ for a symmetry distinct from
lepton number. We accordingly call such theories `charged' majoron models
(CMM's), and denote the associated decay by $\bbcm$.\foot\fttwo{A variation on
this theme in which this broken symmetry is gauged has been discussed in
\ref\carone{C.D.~Carone, Harvard preprint HUTP-93-A007 (1993).}
ref.~\carone.}

Our main results, briefly summarized in ref.~\ourletter, are:

\topic{1} The two classes of models predict {different} electron spectra for
majoron-emitting double beta decay, which may therefore be used to identify the
type of process that is being observed.\foot\ftthree{A similar spectrum
can arise for OMM's if two majorons are emitted simultaneously, as in the models
\ref\twomajoron{R. Mohapatra and Takasugi,\plb{211}{88}{192}.}  
of ref.~\twomajoron.}

\topic{2} If the majorons are electroweak singlets and the couplings
are renormalizable, then $\bbm$ is observable only if there is a neutrino which
mixes appreciably with the electron neutrino and whose mass is at least $\sim
(50 -100)$ MeV.  CMM's are further constrained to have the mass of this neutrino
also not much {\it  heavier} than a few hundred MeV. 
\endtopic

We start, in the following section, with a brief summary of the experimental
situation, parametrizing the size of the effect in terms of the strength of a
hypothetical Yukawa coupling between the majoron and the electron neutrino.
There follows a formulation of the naturalness problem faced by OMM's. This
motivates the introduction and definition of our alternative: the charged
majoron.  

Section (3) proceeds with an analysis of the $\bb$-decay rate for a theory with
generic neutrino masses and Yukawa couplings.  We derive the shape of the
predicted electron sum-energy spectrum for all of the models of interest, as
well as a general momentum-space parameterization of the relevant nuclear
matrix elements as a sum of six form factors.  General formulae for the $\bb$
decay rates are presented in terms of these form factors, which we also
translate into nuclear matrix elements in the nonrelativistic impulse
approximation for the weak interaction currents.

Sections (4) and (5) then apply the general expressions derived in Section (3)
to specific models of the ordinary and charged majoron type. The properties of
the particle spectrum required for a sufficiently large $\bb$ rate are
determined, and the necessity of a neutrino with mass $M \gsim 100$ MeV is
explained. We show that, for CMM's, $M$ must also not be much heavier than this
scale if the observed anomalous $\bb$ rate is to be accounted for. A similar
conclusion follows on less robust grounds from naturalness considerations for
OMM's.  

Having established what conditions are necessary for producing the observed
decay rate, we turn in Section (6) to a discussion of the constraints these
theories must satisfy to avoid conflict with other experiments. Searches for
heavy neutrinos in the decays $K \to e \nu$ and $\pi \to e \nu$ currently
furnish the most restrictive laboratory limits.  Nucleosynthesis is given
particular attention in this section, since it would rule out the existence of
light scalars that are required in both ordinary and charged theories. We show
how these bounds can be evaded by somewhat complicating the various models. 

\section{General Considerations}

We begin by parametrizing the size of the anomalous effect in the data, and 
expounding the theoretical naturalness issue which provides the biggest
challenge in accounting for the excess events.  For the purposes of model
building, the salient features of the anomalous events are that they are 
above where standard $\bbtn$ decays contribute appreciably, yet
below the endpoint for the decays.  These facts preclude their interpretation
as either $\bbtn$ or the neutrinoless $\bbzn$.  

Another crucial input comes from $e^+e^-$ annihilation at LEP. The precise
measurement of the $Z$-boson width for decay into invisible particles
constrains its couplings to putative light scalars.  Any model in which the
rate for $Z \to $ (light scalars) is appreciable, for example that of 
Gelmini and Roncadelli \GR, is ruled out.  We therefore focus on scalars that
are electroweak singlets \CMP.  Although it is possible for majorons to be an
admixture of both singlets and fields carrying electroweak charges, they have
no advantages over purely singlet majorons with respect to the beta decay
anomalies.  In fact these models suffer even more severely from the naturalness
problems outlined later in this section, and so we will not consider them
further.   

\subsection{The Size of the Effect}

There are currently four experiments measuring double beta decay with
sufficient precision to potentially see the excess events observed by the UC
Irvine group. Two report no excess, with one of these quoting an upper bound
\ref\heidelberg{M. Beck \etal, \prl{70}{93}{2853}.}
\heidelberg\ that is marginally in conflict with the Irvine result. 

To compare the effect in various nuclei, we follow the experimental practice of
quantifying the $\bbm$ rate using a hypothetical direct Yukawa coupling between
the electron neutrino and a massless scalar, $\varphi$.  The rate for majoron
emitting double beta decay follows from the Feynman graph of Fig.~(1), evaluated
using the effective interaction 
\label\phenyuk
\eq
\Scl_{\rm phen} = {i\over 2} \geff \; \bar{\nu}_e \, \gamma_5 \, \nu_e
\; \varphi.
\eeq

Table 1 lists the coupling strength, $\geff$, needed to produce  the observed
signals in the various double beta decay experiments. Our analysis used the
nuclear matrix elements ($\Scm = \Avg{[\sn\cdot\sigm - (\gv/\ga)^2]h(r)}$) found
in Staudt \etal\ 
\ref\Staudt{Staudt, Muto and Klapdor-Kleingrothaus, Europhys.~Lett., {\bf 13}
(1990) 31}
\Staudt\ to estimate the rates for the two-neutrino and majoron  decay
modes. The details of how these matrix elements arise are explained more fully
in later sections.  Here $\gv$ and $\ga$ are the axial and vector couplings of
the weak currents to the nucleon and $h(r)$ is a neutrino potential. To quantify
the number of excess events, we choose (by eye) a threshold energy, $E_{\rm
th}$, above which the anomalous events begin and the contribution from ordinary
$\bb$ decay is negligible. The data are taken from ref.~\Moe\ for the elements
\Se, \Mo\ and \Nd, and from the published spectrum of \Ge\ in ref.~\ge. In all
of these cases the excess events comprise $R = 2$ to 3 \% of the total number
observed. Interestingly, $g_{\rm eff}$ lies in the range $\SN{8}{-5}$ to
$\SN{4}{-4}$ for all elements. 

Although the coupling apparently needed for \Mo\ looks disturbingly large
compared to the others, this may be due to uncertainties in the evaluation of
the nuclear matrix elements.  As described in ref.~\Staudt, the $0\nu$ matrix
element for \Mo\ in particular suffers from the near collapse of the random
phase approximation that was employed.  

We also quote here, for comparison, the results of the Heidelberg-Moscow-Gran
Sasso group, who claim a 90\% c.l.~upper bound for \Ge\ of $\geff <
1.8\pwr{-4}$ in \heidelberg. A similar bound of $\geff < 2.0\pwr{-4}$ is
reported for Xe decay by the Neuchatel-SIN-Caltech collaboration
\ref\neuchatel{J.-L.~Vuilleumier \etal, {\it Nucl. Phys.} (Proc. Suppl.) {\bf
B31} (1993).} 
\neuchatel.

\midinsert
$$\vbox{\tabskip=0pt \offinterlineskip
\halign to \hsize{\strut#& #\tabskip 1em plus 2em minus .5em&
\hfil#\hfil &#& \hfil#\hfil &#& \hfil#\hfil &#& 
\hfil#\hfil &#& \hfil#\hfil &#\tabskip=0pt\cr
\noalign{\hrule}\noalign{\smallskip}\noalign{\hrule}\noalign{\medskip}
&& Element && $T^{-1}_{1/2}$(y$^{-1})$ && $R$ && $E_{\rm th}$ && 
$g_{\rm eff}$ &\cr  
\noalign{\medskip}\noalign{\hrule}\noalign{\medskip}
&&\Ge\ && \SNt{2}{-23} && 0.02 && 1.5 && \SNt{1}{-4} &\cr
&&\Se\ && \SNt{2}{-22} && 0.03 && 2.2 && \SNt{8}{-5} &\cr
&&\Mo\ && \SNt{3}{-21} && 0.03 && 1.9 && \SNt{4}{-4} &\cr
&&\Nd\ && \SNt{3}{-20} && 0.02 && 2.2 && \SNt{2}{-4} &\cr
\noalign{\medskip}\noalign{\hrule}\noalign{\smallskip}\noalign{\hrule}
}}$$

\noindent {\eightrm\baselineskip=0.6\baselineskip Table 1: The parameters
required for emission of ordinary majorons in double beta decay. $\ss
T_{1/2}^{-1}$ is the inverse half-life of the anomalous events; and $\ss R$ is
the ratio of anomalous to the total number of events. $\ss E_{\rm th}$ (MeV)
denotes our choice for the threshold value of the sum of the electron energies,
above which essentially only excess events appear. $\geff$ is the
phenomenological coupling (defined in eq.~(\phenyuk) required to explain the
excess rate.} 
\endinsert

Besides these laboratory experiments in which the electron energy spectrum is
directly measured, there are also several geo/radiochemical experiments.
In these the final abundance of daughter products is
measured, so only the total decay rate can be determined. Since the energy
spectrum is unknown, it is impossible to directly determine which process is
responsible for the decay. For comparison with Table 1, we show in Table 2 the
couplings, $\geff$, of eq.~\phenyuk\ that would be allowed assuming the {\it
total} decay rate were due to the majoron-emitting process. Since these values
are comparable with those in Table 1, confirmation of the laboratory excess
events would likely imply a significant role for majoron emission in the
geophysical observations.

The predictions for \Te{}\ are of particular interest because of a recent
measurement of the ratio of decay rates $\zeta \equiv \G$(\Te{130}) $/\G$
(\Te{128}) $ = (2.41\pm 0.06)\times 10^3$.\foot\ftfour{This
\ref\Bern{T.~Bernatowicz \etal, \prl{69}{92}{2341}.} 
result of Bernatowicz \etal\ \Bern\ was used to constrain $\bbom$ emission by
W.~Haxton at Neutrino 92, Granada, Spain  
\ref\Haxton{W.~Haxton, {\it Nucl. Phys.} (Proc. Suppl.) {\bf B31} (1993).}
\Haxton.}  Taking the ratio of the lifetimes is useful because some of
the uncertainties in their experimental determination are expected to cancel. 
As we will explain in more detail in subsequent sections, the significance of
this ratio lies in its strong dependence on the relative phase space for the two
decays \Haxton. It is therefore sensitive to the integrated electron spectrum,
which can discriminate between the different possible decay processes.  

The allowed coupling for \U\ is included here for completeness, although we
have been informed that the discrepancy between the \U\ observations of
Turkevich \etal\ 
\ref\Turkevich{Turkevich, Economou and Cowan, \prl{67}{91} 3211}
\Turkevich\ and the calculations of Staudt \etal\ \Staudt\ have now been
resolved by improving the theoretical estimates.

\midinsert
$$\vbox{\tabskip=0pt \offinterlineskip
\halign to \hsize{\strut#& #\tabskip 1em plus 2em minus .5em&
\hfil#\hfil &#& \hfil#\hfil &#& \hfil#\hfil &#& 
\hfil#\hfil &#\tabskip=0pt\cr
\noalign{\hrule}\noalign{\smallskip}\noalign{\hrule}\noalign{\medskip}
&& Element && $T^{-1}_{1/2}$(y$^{-1})$ && $\Omega$ && $g_{\rm eff}$ &\cr 
\noalign{\medskip}\noalign{\hrule}\noalign{\medskip}
&&\Te{128}\ &&\SNt{1}{-25} && 0.23 && \SNt{3}{-5} &\cr
&&\Te{130}\ &&\SNt{4}{-22} && 30 && \SNt{6}{-5} &\cr
&&\U\ && \SNt{5}{-22} && 33 && \SNt{2}{-4} &\cr
\noalign{\medskip}\noalign{\hrule}\noalign{\smallskip}\noalign{\hrule}
}}$$

\noindent {\eightrm Table 2: The parameters consistent with emission of
ordinary majorons in double beta decay. $\ss T_{1/2}^{-1}$ is the total inverse
half-life, assumed to consist completely of {\it anomalous} events. $\ss\Omega$
is the total phase space available for each decay measured in units of $m_e^7$.
As in Table 1, $\geff$ is the required coupling, defined in eq.~\phenyuk.  The
changes in \Te{}\ relative to ref.~\ourletter\ reflect the new measurements of
ref.~\Bern.} 
\endinsert

\vfill\eject
\subsection{The Naturalness Issue}

One of the first puzzles that must be addressed by any theory of the anomalous
events is how $\bbm$ could be seen without evidence for the neutrinoless
decay, $\bbzn$. If the emitted scalar is the \ngb\ for spontaneous breaking of
lepton number, as in OMM's, then $\bbzn$ must exist at some level due to the
generation of Majorana neutrino masses. 
We argue that OMM's answer this question by requiring some
dimensionful parameter in the scalar potential to be of order 10 keV. 

The small scale arises because the same {VEV}, $u$, that breaks lepton
number in these models typically also generates a Majorana mass for the
electron neutrino whose size is 
\label\Majoranamass
\eq  m_{\nu_e\nu_e} \simeq \geff \; u. \eeq
The Majorona mass gives rise $\bbzn$ decays which would have been seen if it
exceeded the experimental limit
\label\bbznbound
\eq  m_{\nu_e\nu_e} \lsim  1 \eV.  \eeq
Together with the inferred coupling strength, $\geff \sim 10^{-4}$, this bound
implies an upper limit for the lepton-number breaking {VEV} of
\label\vevbound 
\eq u \lsim 10 \keV.  \eeq

One might try to avoid such an artificially small scale simply by having no
breaking at all, $u = 0$. In this case $\bbzn$ is completely forbidden by
lepton number conservation. The question then becomes why the emitted
scalar in $\bbm$ should be so light. Since the experiments resolve events
within 100 keV of the endpoint, the scalar must be no heavier than 100 keV. 
Spontaneous breaking of lepton number naturally satisfies this constraint since
the Majoron is an exactly massless \ngb, but if lepton number is unbroken, the
smallness of the mass would seem to require fine tuning of parameters in the
Lagrangian. 

In either case --- by kinematics if lepton number is unbroken, or from
eq.~\Majoranamass\ if it is broken --- we are led to a mass scale in the scalar
sector of the order of $10 - 100$ keV. Introducing it by hand is  at best
repugnant.  Naturalness demands that the smallness of this new scale, relative
to the higgs VEV, for instance, must be stable under renormalization.
Otherwise we have a new hierarchy problem, which is particularly severe if the
light scalars carry electroweak quantum numbers, as in the triplet majoron
model \GR. In that case, loops involving the electroweak gauge bosons generate
contributions to the scalar potential that are of order $\sqrt{\alpha/ 4\pi} \;
M_W \gsim 100  \MeV$.  

It has been claimed in ref.~\berezhiani\ that a majoron coupling  $\geff \sim
10^{-4}$ is small enough to  generally allow such a hierarchy below the weak
scale to be stable. But in the OMM that these authors consider, the effective
coupling measured in $\bbm$ decay is $\geff \sim g \theta^2$, where the mixing
angle $\theta$ is bounded by neutrino oscillation and decay experiments to be
very small.  This means that the coupling dominating radiative corrections, $g$
rather than $\geff$, is {\it not} small: $g\sim O(1)$. Therefore the
corrections to the small 10 keV scale will tend to be at least four orders of
magnitude bigger than the scale itself and fine-tuning must be invoked. 

We show in Section (4) that the scalar hierarchies in these models {\it
can} be made stable under renormalization by taking advantage of the small
couplings and masses within the neutrino sector, but only in some corners of
parameter space having potentially troublesome phenomenology. For example, the
OMM model of Section (4) points to heavy neutrinos in the mass range of several
hundred MeV that mix appreciably with $\nu_e$. Even though such models are
technically natural, they suffer from the aesthetic problem of requiring
mysteriously small dimensionless scalar self-couplings, $\xi \lsim 10^{-14}$. 

\subsection{Introducing Charged Majoron Models}

The above comparison suggests a third option in which the light scalar mass and
the absence of the neutrinoless decay, $\bbzn$, can { both} be naturally
understood.  To do so, we still assume that the emitted scalar is a
\ngb\ in order to insure its small mass. The absence of $\bbzn$ is also
guaranteed if the spontaneously broken global symmetry is {\it not} lepton
number, which we assume remains conserved.  Thus $\bbzn$ is completely
forbidden because it is a $\Delta L = 2$ process.  The majoron-emitting decay
is still permitted, however, provided that the massless \ngb\ itself carries
lepton charge $L = -2$.   We dub such particles ``charged majorons,'' and show
as one of our main results that they lead to qualitatively different features
for double beta decay, thus allowing them to be distinguished from ordinary
majorons.

\section{General Properties of the Double-Beta Decay Rate}

Next we derive expressions for the rates of the various possible double-beta
decay processes. Although a number of excellent reviews exist 
\ref\DoiTomoda{M.~Doi, T.~Kotani and E.~Takasugi, Prog.~Theor.~Phys.~Suppl.~{\bf
83} (1985) 1; T.~Tomoda, Rept.~Prog.~Phys.~{\bf 54} (1991) 53.}
\ref\reviews{See also: W.C.~Haxton and G.J.~Stevenson, Progress in Particle and
Nuclear Physics {\bf 12} (1984) 409; T.~Tomoda, A.~Faessler and K.W.~Schmid,
Nucl.~Phys. {\bf A452} (1986) 591; S.P.~Rosen, Arlington preprint UTAPHY-HEP-4
(1992); and references therein.}
\DoiTomoda, \reviews, detailed formulae are presented here for several reasons.
Our first goal is to highlight the differences in predictions between OMM's and
CMM's, since the CMM's have not been considered in earlier work.  Secondly we
want to isolate the dependence of our results on the nuclear matrix elements,
since these are the most uncertain factors. For generality, we introduce a
form-factor parametrization of the decay rate which relies simply on the
symmetries of the problem.  Expressions for these form factors in the familiar
nonrelativistic impulse approximation are subsequently derived. 

There are essentially two properties of double-beta decay that can be measured
or computed: the {shape} of the spectrum as a function of the energies of the
two emitted electrons, and the {overall normalization} of this spectrum, which
determines the total decay rate. Only the second of these quantities depends on
the size of the nuclear matrix elements.

Consider the differential decay rate for the four processes to which the
experiments are potentially sensitive: $\bbtn$, $\bbzn$, $\bbom$ and $\bbcm$. 
The amplitudes for the first two depend on the Feynman graphs of Fig.~(2) or
Fig.~(3), respectively.  Those for the majoron emitting processes require
instead the evaluation of Fig.~(1) using the appropriate majoron couplings
(more about which later). 

It is convenient to write the resulting rates as
\label\genericrate 
\eq 
d\G(\bb) = {(\GF\cos\theta_{\sss C})^4 \over 4\pi^3} \; \left| \Sca(\bb)
\right|^2 d\Omega(\bb),
\eeq
where $\GF$ is the Fermi constant, $\theta_{\sss C}$ the Cabibbo angle,
$\Sca(\bb)$ a nuclear matrix element, and $d\Omega(\bb)$ the differential phase
space for the particular process. The observables are taken to be the energies
of the two outgoing electrons, $\epsilon_k$ ($k=1,2)$. Deriving explicit
formulae for $\Sca(\bb)$ and $d\Omega(\bb)$ is the goal of the remainder of
this section.   

Eq.~\genericrate\ shows that the decay rate depends on the nuclear matrix
elements only as an overall multiplicative constant. The only approximation
that must be made to derive this form from the graphs of Figs. (1--3) is the
neglect of the dependence of these matrix elements on the final-state lepton
energies and momenta. This is a good approximation  for the neutrinoless modes
in which we are interested because the final leptons (plus majoron) can carry
at most the endpoint energy, $Q\sim (1-3) \MeV$, while the nuclear matrix
elements are characterized by the nucleon Fermi momentum, $\pf \sim 100 \MeV$.
Corrections to this approximation thus introduce a relative error of order
$Q/\pf\sim$ a few per cent.

\subsection{The Electron Energy Spectrum}

Consider first the electron energy spectrum, $d\Omega(\bb)$ in 
eq.~\genericrate.  This factor is determined solely by the leptonic part of the
appropriate Feynman graph. From Figs.~(1--3) it is straightforward to find the
following results. 

The $\bbzn$ decay is essentially two-body since the nucleus is too heavy to
carry away any appreciable kinetic energy.  The electron phase space is
\label\bbznphsp
\eq  d\Omega(\bbzn) = {1\over 64\pi^2} \; \delta(Q - \epsilon_1-\epsilon_2) 
\prod_{k=1}^2 p_k \epsilon_k F(\epsilon_k) \; d\epsilon_k.
\eeq
Here $p_k = |\bfp_k|$ is the magnitude of the electron
three-momentum, and $Q$ is the endpoint energy for the electron spectrum,
determined by the initial and final nuclear energy levels, $M$ and $M'$, to be
$Q = M-M' - 2m_e$.  $F(\epsilon)$ is the Fermi function, normalized to unity in
the limit of vanishing nuclear charge. 

In contrast, the phase space for the other three processes can be written
in a similar form,
\label\phsp
\eq d\Omega(\bb_i) = {1\over 64\pi^2} \; (Q - \epsilon_1-\epsilon_2)^{n_i}
   \prod_{k=1}^2 p_k \epsilon_k F(\epsilon_k) \;  d\epsilon_k.
\eeq
Only the spectral index $n_i$ differs between $\bbtn$, $\bbom$ and $\bbcm$
decays,
\label\nforeach
\eq  n_{2\nu} = 5; \qquad n_{cm} = 3; \qquad n_{om} = 1.  \eeq
For $\bbtn$ and $\bbom$ these values of $n_i$ simply reflect the phase space
for the corresponding process.  But for $\bbcm$ there are two extra powers of
$(Q -\epsilon_1 - \epsilon_2)$ due to the proportionality of the leptonic
matrix element to the majoron energy, a distinctive and generic feature of
CMM's that we elucidate in Section (3.5) below.  We have assumed that the boson
emitted in $\bbom$ or $\bbcm$ was massless; if it has mass $m$ one must use
$((Q -\epsilon_1 - \epsilon_2)^2 - m^2)^{1/2}$ in place of $(Q -\epsilon_1 -
\epsilon_2)$ in eq.~\phsp.

The difference between $n_{2\nu}=5$ and $n_{om}=1$ has long been
recognized as a way for experimenters to recognize a possible admixture of
these types of decays; they lead to differently shaped curves for the
differential rate, $d\Gamma/d\epsilon$, as a function of the sum of the
electron energies, $\epsilon = \epsilon_1 + \epsilon_2$. The surprising fact
that charged majorons have an index $n_{\sss cM}=3$ intermediate between
$\bbtn$ and $\bbom$ therefore makes it possible, in principle, to determine
whether a distortion in the $\bbtn$ spectrum is due to ordinary or charged
majoron emission.  Fig.~(4) shows the shape of the sum-energy spectra for the
three possible values of the spectral index. 

The spectral shape can also have implications for the total decay rate which,
being an integral over the sum energy spectrum, depends strongly on $n_i$.
Roughly speaking, each successive power of $(Q-\epsilon)$ in
$d\Gamma/d\epsilon$ suppresses the total rate by an additional power
of $Q/(100{\rm\ MeV})$. Therefore  geophysically-determined decay rates, such
as the ratio $\zeta = \G$(\Te{130})$/\G$(\Te{128}) defined in the previous
section, may ultimately prove useful for distinguishing between different
models. Once the relative strength of $\bbtn$ to $\bbm$ decays is better
determined, a definite prediction for $\zeta$ will become possible. If, for
example, the decay rate is dominated by the majoron emitting process, then
$\bbom$ decay predicts too small a ratio \Haxton; we find that $\zeta(\bbom) =
93$. This number includes a factor of 5/7 due to the ratio of nuclear matrix
elements as computed by ref.~\Staudt, and the more significant factor of
(30.4/0.23), due to the difference in phase space for the two decays (see Table
2). Because of the small endpoint energy for \Te{128} compared to that of
\Te{130}, the same ratio for $\bbcm$ decay is much larger: $\zeta(\bbcm) =
770$, and is closer to the experimental value. 

\subsection{The Nuclear Form Factors}

The other observable constraining models of majoron-emitting double-beta decays
is the total rate for any given decay. This requires a knowledge of the matrix
element denoted $\Sca(\bb)$ in eq.~\genericrate, forcing us to deal with the
uncertainties in calculating nuclear transition amplitudes.   The latter can be
written as a sum of six form factors, with which we parametrize the dependence
on nuclear physics.  The form factors can subsequently be expressed (as we do
below) within the context of a given nuclear model.  We start by defining the
form factors, and then use them to specify $\Sca(\bb)$ for the various decay
processes in Sections (3.3) through (3.6) below.  

The nuclear matrix element that appears in the evaluation of Figs. (1--3) is
\label\matrixelement
\eq W_{\alpha\beta}(P,P',p) \equiv (2 \pi)^3 \, \sqrt{ { E E'\over M M'}} \;
\int d^4x \;\bra N'|T^* \left[ J_\alpha(x) J_\beta(0) \right] | N\ket \; 
e^{ipx}.\eeq 
Here $J_\mu = \bar{u} \gamma_\mu (1+\gamma_5) d$ is the weak charged
current that causes transitions from neutrons to protons, and $|N \ket$ and
$|N' \ket$ represent the initial and final $0^+$ nuclei in the decay. $E$ and
$M$ are the energy and mass of the initial nucleus, $N$, while $E'$ and $M'$ are
the corresponding properties for the final nucleus, $N'$. The prefactor,
$\sqrt{EE'/MM'}$, is required to ensure that $W_{\alpha\beta}$ transform as a
tensor since, as is common in the literature, we use nuclear states which are
not covariantly normalized: $\bra \bfp | \bfp' \ket = \delta^3(\bfp - \bfp')$.
The $(2 \pi)^3$ is conventional, and is required in order to put our matrix
elements into the standard form once the overall centre-of-mass motion of the
nucleus is separated out.  

{\it A priori} the tensor $W_{\alpha\beta}$ is a function of the four-momenta,
$P_\mu$ and $P'_\mu$ of the initial and final nuclei, as well as four-momentum
transfer between the two currents, $p_\mu$. This dependence can be
significantly simplified, however.  For $\bbzn$ and $\bbm$, $p_\mu$ is of the
order of the nuclear Fermi momentum $\pf\sim 100$ MeV, whereas the difference
$(P-P')_\mu$ is only a few MeV and may therefore be neglected compared to
$p_\mu$.  Then the dependence of $W_{\alpha\beta}$ on $P_\mu$ and $P'_\mu$ may
be replaced with the single variable $u_\mu$, the common four-velocity of the
initial and final nuclei.  For $\bbtn$, the momentum transfer, $p_\mu$, is
itself also of order the energy released in the decay, and so in this case
$W_{\alpha\beta}$ may be simplified even further by approximating $p_\mu\approx
0$.   

It is also straightforward to show that the Bose statistics of the weak
currents, $J_\alpha$, imply that $W_{\alpha\beta}(u,p) =
W_{\beta\alpha}(u,-p)$. Using the aforementioned approximation, the most
general possible form for $W_{\alpha\beta}$ is \ourletter:\foot\newft{ The
reader should be advised that we define our form factors here differently 
than in ref.~\ourletter.} 
\label\formfactors
\eqa W_{\alpha\beta}(u,p) &= w_1 \; \eta_{\alpha\beta} + w_2 \; u_\alpha
        u_\beta +w_3 \; p_\alpha p_\beta + w_4 \; (p_\alpha u_\beta +
                                        p_\beta u_\alpha) \eolnn
    &+ w_5 \; (p_\alpha u_\beta - p_\beta u_\alpha) + 
     iw_6 \; \epsilon_{\alpha\beta\sigma\rho}
     u^\sigma p^\rho , \eeol \eeq
where the six Lorentz-invariant form factors, $w_a = w_a(u\cdot p, p^2)$, are
functions of the two independent invariants that can be constructed from
$p_\mu$ and $u_\mu$. Under the reflection $p \to -p$, all the $w_i$ are  even
except for $w_4$, which is odd.   

By evaluating the leptonic parts of the $\bb$ matrix elements and contracting
with $W_{\alpha\beta}$, one can show that, to leading order in lepton energies,
only its trace, $W^{\!\phantom{\alpha}\alpha}_\alpha$, enters into the rates for
$\bbtn$, $\bbzn$, and $\bbom$. In terms of the Gamow-Teller and Fermi nuclear
form factors --- which we define in the nuclear rest frame by $\wf = W_{00}$ and
$\wgt = \sum_i W_{ii}$ --- we therefore retrieve the familiar linear combination
\label\trace
\eq W_{\alpha}^{\!\phantom{\alpha}\alpha} = \wf-\wgt. \eeq
For $\bbtn$ we may to a good approximation neglect $p_\mu$. This permits two
important simplifications: ($i$) we may drop all but the form factors $w_1$ and
$w_2$, and ($ii$) we may approximate these two form factors by constants,
$  w_i(u \cdot p, p^2) \simeq w_i(0,0)$. In this limit there is a direct
relation between $w_1$ and $w_2$ with $\wf$ and $\wgt$, given by $w_1
\simeq \nth{3} \wgt$ and $w_2 \simeq \wf + \nth{3} \wgt$.  For $\bbzn$ and
$\bbom$ however, $p_\mu$ is large and so $w_3$ and $w_4$ may also contribute
significantly to $\wgt$ and $\wf$.

The next step is to express $\Sca(\bb)$, and hence the double-beta decay rate
$d\Gamma(\bb)$, in terms of the form factors, $w_a$. Before doing so, we pause
to present explicit expressions for these form factors, modeling the nuclear
decay as the independent decay of its constituent nonrelativistic nucleons.
Besides giving some intuition as to the potential sizes to be expected for
these form factors, these expressions allow a connection between our
form-factor analysis and the nuclear matrix elements that appear in the
literature. 

\vfill\eject
\subsection{The Form Factors in the Nonrelativistic Impulse Approximation}

The common practice in the literature is to provide expressions for the
double-beta decay rates with the nuclear matrix elements computed using 
explicit models of the nucleus. In this section we present expressions for the
form factors using such a model. This gives a point of contact between the
formalism we present here and the rest of the literature. Besides providing a
check on our calculations, the expressions we obtain give some indication
of the size that might be expected for each of the form factors. 

Our evaluation starts by inserting a complete set of states, $| X \ket \bra X
|$, into the matrix element of eq.~\matrixelement. Working in the nuclear rest
frame ( where $EE' / MM' =1$ ) and writing out the time ordering, we have: 
\eqa
W_{\alpha \beta} &= (2 \pi)^3 \int d^4x \sum_X  e^{ipx} \Bigl[ \bra N'|
J_\alpha(x) | X \ket \bra X | J_\beta(0) | N\ket \; \theta(x_0) \eolnn
& \qquad \qquad \qquad \qquad + \bra N'| J_\beta(0) | X \ket \bra X |
J_\alpha(x) | N\ket \; \theta(-x_0) \Bigr] \eolnn
&= -i (2 \pi)^3 \int d^3 \bfx \sum_X  e^{ - i \bfp \cdot \bfx} \left[ 
{ \bra N'| J_\alpha(\bfx) | X \ket \bra X | J_\beta(0) | N\ket \over
p_0 + E_{\sss X} - M'+i\varepsilon} \right. \eolnn 
& \qquad \qquad \qquad \qquad \left. - \; { \bra N'| J_\beta(0) | X \ket \bra X
| J_\alpha(\bfx) | N\ket \over p_0 - E_{\sss X} + M-i\varepsilon} \right]. 
\eeol \eeq 
Contact with the literature can be made once we perform the following 
approximations: 
\topic{(1) The Closure Approximation} 
In this approximation a sum over intermediate states of the form $\sum_X F(E_X)
|X\ket\bra X|$ is simplified by replacing the $X$-dependent prefactor,
$F(E_X)$, by $F(\ol{E})$ where $\ol{E}$ is the energy averaged over the states
that contribute to the matrix element in question. In the present example we
may also use the information that $M - M'$ is much less than $M$ and $\ol{E}$ to
replace $M'$ with $M$ throughout. 
\topic{(2) The Nonrelativistic Impulse Approximation} 
The next simplification is to model the nuclear decay in terms of the
independent decay of its constituent nucleons, which are taken to be
nonrelativistic. We work in the position representation, as is conventional in
nuclear physics. In this representation, the weak currents acting on the
constituent nucleons takes the following form: 
\label\impulse
\eqa J_0(\bfx) = & \sum_n \delta( \bfx - \bfr_n)
        \tau_n^+ \left( \gv - \ga C_n \right) + O(v^2/c^2) \eolnn
     \Bfj( \bfx) = & \sum_n \delta(\bfx - \bfr_n) 
        \tau_n^+\left( \ga \vec\sigma_n - \gv \Bfd_n
        \right) + O(v^2/c^2), \eeol\eeq
where we have included terms up to $O(v/c)$ in the nucleon velocities. The sum
here runs over the constituent nucleons, with the position of the `$n$th'
nucleon denoted by $\bfr_n$. The operator $\vec \sigma_n$ similarly
denotes the Pauli spin matrices acting on the $n$th nucleon spin, while
$\tau_n^+$ is the isospin raising operator for this nucleon. As in Section
(2.1), $\gv \simeq 1$ and $\ga \simeq 1.25$ represent the usual vector and axial
couplings of the nucleon to the weak currents.

The operators $C_n$ and $\Bfd_n$ represent the $O(v/c)$ contributions to the
weak currents, and are included here since some of the form factors vanish in
the limit that $v = 0$. They are defined in terms of the initial and final
four-momenta of the decaying nucleon, $(E_n,\Bfp_n)$ and $(E_n',\Bfp_n')$, the
Pauli spin-matrices, $\vec \sigma_n$, the mass of the pion, $m_\pi$, 
and the mass of the proton, $M_p$, by \DoiTomoda,  
\eqa C_n &= (\Bfp_n + \Bfp'_n) \cdot \vec\sigma_n /(2M_p) 
- (E_n-E_n')(\Bfp_n-\Bfp'_n)\cdot\vec\sigma_n/m^2_\pi, \eolnn
\Bfd_n &= \left[ (\Bfp_n + \Bfp'_n) + i \mu_\beta (\Bfp_n - \Bfp'_n) \times
\vec\sigma_n \right]/(2M_p).  \eeol \eeq 
Here $\mu_\beta = \hf \,( g_p - g_n) \simeq 4.7$ is a combination of the proton
and neutron spin $g$-factors that originates from the contribution of `weak
magnetism.' 
\endtopic

The final step is to separate the overall motion of the nucleon centre-of-mass,
$\Bfr$, out of the nuclear wavefunction. For a nucleus labelled by its overall
momentum, $\Bfp$, as well as its other quantum numbers, $a$, we write: 
\eq
\bra \bfr_1,\dots,\bfr_{\sss A}| \Bfp, a \ket \equiv { e^{i \Bfp \cdot \Bfr}
\over (2 \pi)^{3/2}} \; \bra \hat\bfr_1, \dots, \hat\bfr_{\sss A} | a \dket,
\eeq
where the `reduced' coordinates, $\hat\bfr_n$, are subject to the constraint
$\sum_n \hat\bfr_n = 0$. 

These approximations give the following results for $\wf$ and $\wgt$:
\label\connection
\eqa \wf &= {2i \mu \gv^2\over p_0^2 - \mu^2 + i\varepsilon}
\; \dbra N'|\sum_{nm}e^{-i\bfp\cdot\bfrnm}\tau^+_n\tau^+_m |N \dket;\eolnn
         \wgt &= {2i \mu \ga^2\over p_0^2 - \mu^2 + i\varepsilon} \;
\dbra N'|\sum_{nm} e^{-i\bfp \cdot \bfrnm} \tau^+_n\tau^+_m 
        \; \vec\sigma_n \! \cdot\! \vec\sigma_m |N \dket, \eeol \eeq
where $\mu \equiv \ol{E} - M$ is the average excitation energy of the
intermediate nuclear state, and $\bfrnm$ is the separation in position between
the two decaying nucleons. We neglect the $O(v/c)$ corrections to this
expression. 

The only other combination of form factors which arise for $\bbtn$, $\bbzn$,
$\bbom$ and $\bbcm$ decays are $w_5$ and $w_6$, and these arise only in $\bbcm$.
In the impulse approximation we are using, these expressions vanish at lowest
order in $v/c$, forcing us to go to the next higher order. We find that
\label\wfiveandsix
\eqa w_5 &= {i\mu \bfp\over |\bfp|^2 (p_0^2 - \mu^2 + i\varepsilon)} \; 
\cdot \dbra N'|e^{-i\bfp\cdot\bfrnm} \left[ \ga^2 (C_n \vec\sigma_m-
        C_m\vec\sigma_n) +\gv^2 (\Bfd_m -\Bfd_n) \right] |N \dket ;\eolnn
     w_6 &= {\mu \ga \gv \bfp \over|\bfp|^2(p_0^2 - \mu^2 + i\varepsilon)}
       \; \cdot \dbra N'| e^{-i\bfp\cdot\bfrnm} \left[ \Bfd_n\times\vec\sigma_m
        + \vec \sigma_n \times \Bfd_m\right] |N \dket, \eeol \eeq
in which $\sum_{mn}\tau_m^+\tau_n^+$ are implicit.

We may now complete the calculation by expressing the various double-beta decay
amplitudes, $\Sca(\bb_i)$, in terms of the nuclear form factors $w_1$ through
$w_6$. For this purpose we must specify the form for the interactions and
neutrino masses to be used in evaluating Figs.~(1--3). We consider each of the
four decay processes separately in the following sections. 

\subsection{The $\bbtn$ Rate}

For completeness we start with the standard two-neutrino decay, $\bbtn$.
Evaluating the total rate using the leptonic part shown in Fig.~(2), and
comparing with eq.~\genericrate, we deduce that the nuclear part of the
amplitude is approximately
\label\bbtnrate
\eqa \Sca(\bbtn) &\approx {2 \over \pi \sqrt{15}} \; \left[{W_\alpha}^\alpha
    \right]_{p_\mu =0}, \eolnn 
   &\approx {2 \over \pi \sqrt{15}} \; \left[ 4 w_1(0,0) -
        w_2(0,0) \right], \eolnn 
   & \approx {2 \over \pi \sqrt{15}} \; \left[ \wf(0,0) - \wgt(0,0) \right]. 
\eeol \eeq 
For simplicity all final lepton energies and masses have been ignored. Thus
only the form factors evaluated at zero argument appear because, for $\bbtn$
decay, conservation of momentum and energy determines the nucleon recoil
four-momentum, $p_\mu$, in terms of the energy-momentum of the final-state
leptons.

\subsection{The $\bbzn$ Rate}

To evaluate Fig.~(3) for the $\bbzn$ decay rate, one must know the neutrino
mass spectrum. We consider a general mass matrix for an arbitrary set of
Majorana neutrinos,
\label\neutrinomasses
\eq  \Scl_{\rm mass} = - \hf \; \bar{\nu}_i \left( m_{ij} \Pl + m^*_{ij}
    \Pr \right) \nu_j, \eeq
where $m_{ij} = m_{ji}$ is the left-handed neutrino mass matrix, and $\Pl$
($\Pr$) are the usual projectors onto left-handed  (right-handed) spinors.
The physical masses $m_i$ are given by the square roots of the eigenvalues of
the matrix $m^\dagger m$ --- { not} necessarily by the eigenvalues of $m$
itself, which may be complex.  The electron-flavor row of the associated
`Kobayashi-Maskawa-type' matrix for the weak charged-current interactions is
denoted by $V_{ei}$. 

With this choice the $\bbzn$ decay matrix element becomes
\label\bbznrate
\eq \Sca(\bbzn) = 8\sqrt{2}\pi \sum_i V_{ei}^2 \; m_i \int {d^4p \over
(2\pi)^4} \;
   \left({W^{\!\phantom{\alpha}\alpha}_\alpha \over p^2 - m_i^2 +
   i\varepsilon}\right). \eeq 
Although the range of integration runs over all possible neutrino four-momenta,
$p_\mu$, the nuclear form factors $w_a$ act to cut the integrals off at the
Fermi momentum and energy, $\pf$ and $\EF$. The contributions from heavy
neutrinos thus become suppressed, decoupling as $1/m$, as $m$ starts to exceed
this scale.

Using the approximations of Section (3.3) for the form factors in
${W_\alpha}^\alpha$ leads to the familiar Gamow-Teller and Fermi expressions, 
\label\znnmes
\eq \int {d^4p \over (2\pi)^4} \;
   \left({W^\alpha_\alpha \over p^2 - m_i^2 +
   i\varepsilon}\right) = {1\over 4\pi}\dbra N' | \; h(\bfrnm ;m_i)
        \left(\gv^2 -\ga^2 \vec\sigma_n \cdot\vec\sigma_m\right) |N\dket, 
\eeq
where $h(\bfrnm ;m_i)$ is the neutrino potential function defined by:
\label\nupotential
\eq h(\bfrnm ;m) = {1\over 2\pi^2} \int d^3p\; {\exp(-i\bfp\cdot\bfrnm) \over
\omega(\omega+\mu)}; \qquad \omega = (p^2+m^2)^{1/2}. \eeq
Again, $\sum_{mn}\tau_m^+\tau_n^+$ is implicit in these expressions.

An important special case is that in which  the neutrino masses are negligible
compared to the nuclear scale, $\pf$. Then one can use the massless propagator
in eq.~\bbznrate\ and make the replacement $\sum_i V_{ei}^2 \; m_i =
m_{\nu_e\nu_e}$, since the integral is to a good approximation independent of
$i$.  Thus the rate vanishes in the absence of a direct Majorana mass for the
electron neutrino, as it should. 

\subsection{The Rate for Ordinary Majoron Emission}

For majoron-emitting decays we wish to evaluate Fig.~(1), and this
requires a knowledge of the neutrino--majoron coupling. For generality's
sake we take the form
\label\yukawa
 \eq \Scl_{\varphi\nu\nu} = - \hf\; \bar{\nu}_i (a_{ij} \Pl  + b_{ij}  \Pr)
   \; \nu_j \; \varphi^* + c.c. \eeq
If the scalar field is real then \yukawa\ still applies, but with the
restriction that $b_{ij} = a^*_{ij}$. For example, the phenomenological
interaction of eq.~\phenyuk\ represents the case of a single neutrino
with $b^*_{\nu_e\nu_e} = a_{\nu_e\nu_e} = -i\geff$.

Evaluating Fig.~(1) using this interaction and neglecting, as before, the final
state lepton energies and momenta leads to the amplitude
\label\bbomrate
 \eq  \Sca(\bbom) = 4\sqrt{2} \sum_{ij} V_{ei} V_{ej}
  \int {d^4p \over (2\pi)^4} \; \left[ { W_\alpha^{\!\phantom{\alpha}\alpha} 
        (a_{ij} m_i m_j + p^2 b_{ij}) \over 
  (p^2 - m_i^2 +i\varepsilon ) (p^2 - m_j^2 +i\varepsilon) } \right] . \eeq
For neutrino masses that are much smaller than $\pf$ this expression
simplifies to the form 
\label\masslessbbomrate
 \eq  \Sca(\bbom) \cong - 4\sqrt{2} \left[ \sum_{ij} V_{ei} V_{ej} b_{ij}
        \right]
  \int {d^4p \over (2\pi)^4} \; \left( { W^{\!\phantom{\alpha}\alpha}_\alpha
  \over p^2 +i\varepsilon } \right), \eeq
which involves the same combination of nuclear matrix elements as appears in
eq.~\bbznrate\ for $\Sca(\bbzn)$ with light neutrinos, a result first pointed
out in ref.~\georgi. The sum over mass eigenstates simply gives the coupling in
the flavor basis, $b_{\nu_e\nu_e}$, which must vanish in a renormalizable
theory if the majoron comes from an electroweak singlet field. Thus in
renormalizable singlet-majoron models it is necessary for at least one neutrino
to have a mass $m\gsim \pf \sim (50 - 100)$ MeV. Such a neutrino can generate
an effective, nonrenormalizable $b_{\nu_e\nu_e}$ coupling upon being integrated
out. Notice that this observation rules out the simplest singlet-majoron model
\CMP, (in which the standard model is supplemented by a singlet scalar
field without additional intermediate-mass neutrino species) as an explanation
for the anomalous $\bb$ events. 

\subsection{The Rate for Charged Majoron Emission}

For charged majorons the interaction \yukawa\ must be further constrained to
reflect the fact that $\varphi$ now carries lepton number.  Suppose that the
global symmetry for which $\varphi$ is the \ngb\ acts on the neutrino
fields in the following way: $\delta \nu = i  (q \, \Pl - q^\transp \, \Pr)
\nu$, with generator  represented by the matrix $q$. Then, as shown in
Appendix A, the majoron coupling matrix to  neutrinos may be written as 
\label\ngbyukawas
\eqa  a &= -{ i\over f} ( q^\transp m + m q ), \eolnn
      b &= +{ i\over f} ( q m^* + m^* q^\transp ),  \eeol   \eeq
where $f$ is the decay constant, proportional to the symmetry-breaking scale. 
Note that \ngb s carrying an unbroken charge are associated with nonhermitian
generators (for example, the longitudinal component of the $W^\pm$ bosons), so
that $q^\transp\neq q^*$ in what follows. 

Eqs.~\ngbyukawas\ are equivalent to the statement that it is possible to
redefine the neutrino fields in such a way as to ensure that the
neutrino--boson coupling has the derivative form
\label\derivative
\eq  \Scl_{\varphi\nu\nu} = {i\over 2f} \; \bar{\nu} \gamma^\mu (q \Pl -
   q^\transp \Pr) \nu
   \; \partial_\mu \varphi + c.c. \eeq
The equivalence of this interaction with the Yukawa formulation is demonstrated
explicitly for double-beta decay in Appendix B.

The big surprise now comes when eqs.~\ngbyukawas\ are substituted into the
result \bbomrate\ for the $\bbom$ decay rate. As is shown by brute force using
the Yukawa couplings in Appendix C, the result vanishes identically! This is a
reflection of the general statement that the amplitude, $\Sca(\bbcm)$, vanishes
as the energy of the emitted majoron goes to zero [recall that we ignored all
final state momenta in deriving \bbomrate], a fact which is most easily seen
using the variables for which the neutrino-majoron coupling takes its
derivative form as in eq.~\derivative. 

This result depends crucially on having the emitted \ngb\ carry an unbroken
quantum number --- in this case lepton number.  The same result does not apply
to ordinary majorons, even if they are true \ngb s rather than being massive. 
This statement may be puzzling on reflection, since in this case also one can
put the majoron-neutrino coupling into the derivative form of eq.~\derivative.
The resolution of the paradox is that for OMM's the rest of the amplitude is
singular in the limit of vanishing majoron energy, leaving a nonzero result. 
For the details of this argument we refer the reader to Appendix D. 

The upshot is that in CMM's one must work to next higher order in the final
lepton energies than was done to get eq.~\bbomrate. The extra factors of the
majoron momentum can be put into $d\Omega(\bbcm)$, and account for the
difference between $n_{cm}$ and $n_{om}$ in eq.~\nforeach. In the rest frame
of the decaying nucleus, the nuclear matrix element turns out to be\foot\ftsix{
We thank C. Carone for pointing out an error (corrected here) in this equation
as it appeared in ref. \ourletter.} 
\label\bbcmamp
   \eq \Sca(\bbcm) = 8\sqrt{2} \sum_{ij} V_{ei} V_{ej} b_{ij}  
   \int {d^4p \over (2\pi)^4} \; \left[ {  \bfp^2 (w_5 + w_6 ) \over  
 (p^2 - m_i^2 +i\varepsilon ) (p^2 - m_j^2 +i\varepsilon)} \right]. \eeq 
Whereas previously it was the trace of $W_{\alpha\beta}$ that arose in the
decay rate, here it is the skew-symmetric part, parameterized by the form
factors $w_5$ and $w_6$, that appear. In the nuclear rest frame these form
factors are given by $w_5 = p_i(W_{0i}-W_{i0})/(2|\bfp|^2)$ and $w_6 =
\epsilon_{ijk}p_iW_{jk}/(2|\bfp|^2)$. 

Using the approximations of Section (3.3) for these form factors leads to the
formulae of ref.~\ourletter.\foot\ftseven{
These two matrix elements correspond to what was called $\ss A_1$ in 
ref.~\ourletter.  We have corrected the erroneous coefficient of $\ss 7/9$ which
multiplies $\ss A_1^2$ there.  There is a $\ss p$-wave contribution to the amplitude 
which we called $\ss A_2$, omitted here because it is expected to be much smaller. 
The amplitude $\ss A_3$ is also omitted here because it can be seen to vanish
identically.} We note that the neutrino potential that results from doing the
momentum integral in eq.~\bbcmamp\ does not give the usual expression,
eq.~\nupotential, because of the different momentum dependence of the form
factors \wfiveandsix.

In fact, there are a number of important differences between the charged majoron
amplitude \bbcmamp\ and the corresponding result for ordinary majorons,
eq.~\bbomrate: 

\topic{1} Eq.~\bbcmamp\ depends on completely different form factors than the
corresponding expression for any other kind of majoron-emitting double beta
decay.  In fact, we know of no variety of $\bb$ which depends on 
$w_5$,\foot\fteight{except for a highly subdominant Coulomb/recoil correction
to $\ss \bbzn$ that would give an $\ss S-P_{1/2}$ final state for the electrons--see 
eq.~(C.2.12b) of reference \DoiTomoda}\ and this matrix element therefore
appears not to have been computed by anyone yet.  On the other hand $w_6$ would
appear in $\bbzn$ if there were right-handed currents, and it has been
calculated in 
\ref\muto{K.~Muto, E.~Bender and H.V.~Klapdor, Z.~Phys.~{\bf A334} (1989) 187.}
ref.~\muto. An interesting feature of this computation is that the value of
$w_6$ does {\it not} appear to be suppressed by the nucleon velocity, $v/c$, as
would have naively been expected, but is instead rather large. 

\topic{2} In the limit where all neutrinos are much lighter than $\pf$, the
flavor dependence of the amplitude becomes proportional to the same
combination of couplings as appeared for OMM's: $\sum_{ij} V_{ei} V_{ej} b_{ij}
= b_{\nu_e\nu_e}$.  This direct coupling to the electron neutrino must vanish
in any renormalizable CMM's, because we have assumed that the \ngb s to all be
electroweak singlets. Analogously to OMM's, it follows that in any CMM at least
one of the neutrinos must have an appreciable mass: $m_i \gsim \pf \sim (50
- 100)$ MeV. 

\topic{3} As may be seen from eq.~\bbcmamp, if either of the neutrinos in the
graph are large compared to $\pf$, then the result becomes suppressed by at
least {\it two} powers of the heavy neutrino mass: $1/m^2$. Notice that this is
a stronger suppression than the $1/m$ behavior that follows from eq.~\bbomrate,
for OMM's. 
\endtopic 

\vfill\eject
\section{Ordinary Majoron Models}

We now construct a viable alternative to the original singlet and triplet
majoron models, since these are not able to yield an observable rate of double
beta decay while still satisfying all other experimental constraints.  In this
section we focus on ordinary majorons, reminding the reader that here
`majoron' means any light scalar with couplings to neutrinos, regardless of
whether it is a \ngb.  It will be shown that if $\bbom$ occurs at the rate
suggested by present experiments, one can infer that the
masses and mixing angles of the neutrinos are in a range where they are
potentially observable by other kinds of experiments. 

\subsection{A Minimal Model}

It is easy to invent a minimal ordinary majoron theory encompassing
the low-energy effects of more complicated physics at the electroweak scale. 
Let $\varphi$ be a complex electroweak-singlet scalar carrying $-2$ units
of lepton number.   The lowest dimensional operator coupling
$\varphi$ to leptons, while respecting the gauge symmetries and global lepton
symmetry, is
\label\bbcoupling
\eq  \Scl = - {\kappa \over 2 M^2} \; (\bar{L}H)(H^\transp L^c) \; \varphi
        + c.c. \eeq
Here $L = {\nu_e \choose e}_{\sss L}$ and $H$ are the usual left-handed-lepton
and Higgs doublets. This interaction can be derived from a more fundamental
theory, such as the one given in the next section, by integrating out heavy
particles of mass $M$. $\kappa$ is a dimensionless number that depends on the
coupling constants of the underlying theory. 

Once the Higgs doublet is replaced by its expectation value, $\Avg{H} =
v = 174$ GeV, eq.~\bbcoupling\ reduces to a coupling of the form of
eq.~\yukawa, with strength
\label\ommcouplingstrength
\eq  a_{\nu_e\nu_e} = 0, \quad b_{\nu_e\nu_e}={\kappa v^2
\over M^2}. \eeq 
The imaginary part of $\varphi$ therefore couples axially as in eq.~\phenyuk,
with $\geff = \kappa v^2/\sqrt{2} M^2$. Because of the requirement $\geff
\simeq 10^{-4}$, it follows that $M/\sqrt{\kappa} \simeq 10$ TeV, consistent
with the assumption that the particles of mass $M$ can be integrated out when
analysing double beta decay. 

If the light scalar, $\varphi$, should also develop a VEV, then the effective
coupling of eq.~\bbcoupling\ also induces an majorana electron-neutrino mass,
$m_{\nu_e} = \geff \, \Avg{\varphi}$, which is consistent with the present upper
bound only if $\Avg{\varphi} \lsim 10$ keV. The simplest assumption is that
$\Avg{\varphi}=0$. This illustrates the general arguments of section (2.2)
in the present example.

The electron spectrum for $\bbom$ that would be predicted by this effective
coupling can be consistent with the excess events that are seen, provided that
at least one scalar mass eigenstate is lighter than 100 keV. This conclusion
holds regardless of whether $\Avg{\varphi}$ is strictly zero or not, since the 
decay rate found using the scalar couplings of eq.~\bbcoupling\ in the general
expression of eq.~\bbomrate\ is sufficiently large even for massless neutrinos,
as would be implied by a vanishing VEV.  

The alert reader may wonder how the above model can lead to observable scalar
emission even when the neutrinos are massless, since this is in apparent
contradiction to the general result for $\bbom$ decay that was stated in Section
(3.5) above. There we claimed that $\bbom$ is suppressed if all neutrinos are
much lighter than the scale, $\pf$, of the nuclear matrix elements.  The
contradiction is only apparent, however, because the argument of Section (3.5)
presupposed only dimension-four (\ie\ renormalizable) Yukawa couplings, and so
does not include those of eq.~\bbcoupling.  In fact, this effective coupling
can be obtained by integrating out the heavy neutrino that is required by the
general arguments in a renormalizable theory, as we demonstrate shortly.

An imperative question in this scenario is why the potential for $\varphi$
should contain such a small scalar mass or vacuum expectation value. But
somewhat surprisingly, the hierarchy between this small scale and the weak
scale {\it is} technically natural in the sense of being stable against
renormalization, at least within the low-energy effective theory below the
heavy scale, $M$.  Quantitatively, there are two types of dangerous terms
within the scalar potential of the effective theory,
\label\dangerousterms
\eq  \rho^2\,\varphi^*\varphi, \qquad \xi \,\varphi^*\varphi
        \, H^\dagger H, \eeq
whose coefficients must be extremely small, $\rho \lsim 10$ keV and $\xi \lsim
10^{-14}$, if $m_\varphi$ is to be kept $\lsim 10$ keV. 
If we choose to define the running of these couplings within the
decoupling-subtraction renormalization scheme,\foot\ftnine{This scheme consists
of the usual $\ol{\ss MS}$ scheme, supplemented by the explicit integrating
out of any heavy particles as the renormalization point is reduced below the
corresponding thresholds 
\ref\DS{A useful review may be found in I.~Hinchliffe, in {\it TASI Lectures in
Elementary Particle Physics}.}
\DS.} then both couplings run logarithmically, except for the discontinuous
quadratic contributions when a particle is integrated out at its threshold. 
The initial conditions for the renormalization-group (RG) equations in this
scheme are given by the values of the couplings, \eg\ $\rho(M)$ and $\xi(M)$,
at the heavy-physics scale $\mu = M$, where the effective theory is matched
onto the underlying theory.  Provided that these initial values are small, the
logarithmic RG evolution through the scales $\mu < M$ in the effective theory
keeps them small. The same is true for the nonlogarithmic contributions
that arise when the $W$ and $Z$ bosons are integrated out, since these
particles couple only very weakly to $\varphi$. Furthermore, even though the
coupling, $\geff$, to light neutrinos is not particularly small, the effects of
the operator \bbcoupling\ in loop diagrams are suppressed within the effective
theory by the small (or vanishing) $\nu_e$ mass.  

Although the small scalar mass is stable within the effective theory below the
scale $\mu=M$, the difficult issue is whether there exists a model for the
physics at $\mu = M$, which can produce the effective coupling, eq.~\bbcoupling,
and still not generate large scalar self couplings. Such a question can only be
addressed within the context of the underlying renormalizable interactions,
which are the subject of the next section.

\subsection{A Renormalizable Model}

It is useful to look for a ``fundamental'' theory whose low-energy limit is the
phenomenological model in the previous section.  One would like to know whether
such a theory exists,  whether it has any additional observable consequences,
and how much fine-tuning it requires. Naturally we seek a candidate with the
smallest number of new particles.  With hindsight, the simplest choice appears
to be the addition of a Dirac neutrino, whose mass will turn out to be in the
range of $\pf\sim (50-100)$ MeV.  Of course we also must include the singlet
scalar that is emitted in $\bb$ decay.   The Dirac neutrino can be described as
two singlet left-handed neutrinos $s_{\sss\pm}$, whose lepton number charges
are $\pm 1$, and the singlet scalar field must have lepton charge $-2$. The
most general renormalizable couplings of the new particles, consistent with the
assumed symmetries, are 
\label\ommrenmodel
\eq \Scl = - \lambda \bar{L}H \Pr \sm - M \bar{s}_+ \Pr \sm
        -\frac12 \gp \bar{s}_+ \Pr \sp \varphi - \frac12 \gm 
        \bar{s}_- \Pr \sm \varphi^* + c.c. \eeq
For simplicity we assume that lepton number is not spontaneously
broken: $\Avg{\varphi} =0$. The spectrum then contains three massless
neutrinos, $\nu_e'$, $\nu_\mu$ and $\nu_\tau$, together with a massive Dirac
neutrino, $\nu_h$.  The relation between the left-handed weak-interaction
eigenstates and the left-handed mass eigenstates is
\label\mixings
\eqa    \nu_e  &= \nu_e' \; \cos\theta + \nu_h \; \sin\theta, \eolnn
        \sm &= \nu_h^c, \eolnn
        \sp &= - \nu_e' \; \sin\theta + \nu_h \; \cos\theta, \eeol \eeq
where $\tan\theta = \lambda v/M$ and  $\nu_h^c$ is the charge conjugate of
$\nu_h$.  For $\nu_h$ masses, $M_h$, in the range of present interest the
universality of leptonic weak interactions requires that $\theta\lsim 0.1$, very
conservatively; thus we have the hierarchy $M / \lambda v \gsim 10$, and $M_h =
\sqrt{M^2 + \lambda^2 v^2} \simeq M$. If $\nu_h$ is very heavy compared to $\pf$
it may be integrated out, resulting in an effective coupling of the form of
eq.~\bbcoupling, with $\kappa / M^2  =  \lambda^2 \gp / M_h^2$.  

There are two light scalars which can be emitted in double-beta decay in this
model, corresponding to the real and imaginary parts of the complex field,
$\varphi$. The total rate for $\bbom$ decay is given by eq.~\masslessbbomrate,
where the Yukawa couplings of the scalar to the light neutrino are   
\label\thecoupling
\eq  a_{\nu_e' \nu_e'} = 0, \qquad b_{\nu_e' \nu_e'} = \sin^2 \theta \gp. \eeq
This latter coupling is also equal to $\geff$, so the experimentally
suggested value of $\geff = 10^{-4}$ may be obtained by varying the parameters of
the renormalizable model in the range $0.01 \lsim \gp \lsim 1$, and $0.01 \lsim
\sin\theta \simeq  \lambda v/M \lsim 0.1$.  Setting $\lambda=1$, we get an upper
limit of $M_h\sim 10$ TeV for the heavy neutrino mass.  But if $\lambda = 
10^{-4}$, for example, then $M_h\sim 100$ MeV, which is the smallest it can be
before the amplitude starts to become suppressed by powers of $M_h/\pf$.  In
that case we must use the more exact expression for the leptonic part of the
matrix element.  This is accomplished by making the replacement 
\label\moreexactly
\eqa {\sum V_{ei}V_{ej}b_{ij}\over p^2 + i\varepsilon }\quad\to\quad &
        \gp\cos^2\!\theta\sin^2\!\theta \left({1\over p^2+i\varepsilon}
        -{2\over p^2-M_h^2+i\varepsilon}+ {p^2\over (p^2-M_h^2+i\varepsilon)^2}
        \right) \eolnn & +
        \gm\sin^2\!\theta {M_h^2 \over (p^2 - M_h^2 +i\varepsilon)^2}\eeol \eeq
in eq.~\masslessbbomrate, which agrees with $b_{\nu_e' \nu_e'} \cos^2 \theta$ in
the limit of large $M_h$. 

\subsection{Naturalness}
\nobreak
Having specified the particle content at the intermediate mass scale, we can
now return to the question of how natural is the smallness of the scalar
masses.  We saw in the previous section that if the initial values $\xi(M)$ and
$\rho^2(M)$ are small, then $\xi(\mu)$ and $\rho(\mu)$ remain small as $\mu$
runs to lower energies, for which the effective lagrangian \ommrenmodel\ is
valid.  But this by itself is not enough; in addition we must establish whether
the matching conditions at the heavy-neutrino threshold $\mu = M$  are
consistent with small values for $\xi(\mu)$ and $\rho(\mu)$ at scales  $\mu >
M$.  We regard the parameters as being naturally small only if no delicate
cancellation is needed between their values above $\mu =M$ and the quadratic
contribution arising when $\nu_h$ is removed from the effective theory. 

The contribution to $\rho(M)$ and $\xi(M)$ due to integrating out the heavy
neutrino is easily estimated from the graphs of Fig.~(5), in which $\nu_h$ is
the virtual particle:  
\label\thresholdterm
\eqa  \delta\rho^2(M)  &\sim  {\gp \gm \over 16 \pi^2} \; M^2, \eolnn
        \delta \xi(M) &\sim {\gm^2 \lambda^2 \over 16 \pi^2}. \eeol \eeq
Contrary to the prejudice that a $\rho =10$ keV scalar mass requires
extreme fine tuning, we see that it is possible to keep $\delta\rho\ltwid
\rho=10$ keV and yet have $\geff=10^{-4}$ using plausible values of the
underlying couplings.  For example  $\gp \simeq 10^{-2}$, $\gm \simeq \lambda
\simeq 10^{-4}$ and $\sin\theta \simeq 0.1$ implies $M_h \simeq 100 \MeV$,
which then implies, from eqs.~\thresholdterm, $\delta\rho^2(M) \simeq (10
\keV)^2$ and $\delta\xi(M) \simeq 10^{-18}$. 

In the above scenario we have $\nu_e$ mixing strongly with a neutrino in the
mass range of several hundred MeV; a choice with potentially strong
phenomenological consequences (see Section (6)). However $M_h$ can be pushed to
much higher values by letting $\gm$ become smaller, since this has no effect
on the effective Majoron coupling, eq.~\thecoupling, that is relevant to $\bb$
decay. While it might seem less pleasing aesthetically to have $\gm\ll \gp$, 
there are no logical grounds for excluding this possibility. In this case the
model is safe from any of the constraints to be discussed in Section (6) that
follow from an MeV scale neutrino.

\section{Charged Majoron Models}

We now repeat the above exercise for charged majoron models, \ie, to construct
the simplest example both as a low energy effective theory and a renormalizable
one.  It will be seen that our CMM has some close similarities to the OMM
just constructed.  In contrast however we will find that a heavy neutrino in
the 100 MeV mass range is not merely suggested, but required, in order to
achieve a high enough $\bbcm$ rate. 

To motivate the specific example, we start with some general considerations.
Consider the spontaneous breaking of a global symmetry group $G$ down
to a subgroup $H$. The resulting \ngb s can carry quantum numbers with respect
to unbroken charges in $H$ only if the original group $G$ is nonabelian, and
the unbroken charges do not all commute with the broken generators of $G$. In
the Standard Model itself, a global nonabelian symmetry acting upon the leptons
is precluded by their Yukawa couplings to the standard Higgs, or equivalently
by the charged-lepton masses.  We must extend the low-energy particle content in
order to devise such a symmetry.  In so doing, it is prudent to let the new
particles be electroweak singlets lest dangerous couplings arise between the
massless \ngb s and charged leptons or electroweak bosons.  

If the new electroweak singlet neutrinos are integrated out, we obtain a
low-energy effective coupling of the charged majoron to light neutrinos. It is
instructive to write down the lowest-dimension such interaction that is possible
since this reveals many of the features that are common to all underlying
models. For CM's the general result that the $\bb$ amplitude must be
proportional to the CM momentum (see Appendix B) suggests using field
variables for which the derivative couplings are explicit. Because the CM 
carries lepton number $L=-2$, the usual interaction of the form $\bar\nu_e
\gamma_5 \gamma_\mu \nu_e \partial_\mu \varphi$ is not allowed; the current has
$L=0$. Rather, we need an even number of gamma matrices, 
\label\cmcoupling
\eq  M^{-4}\bar L H (a_1 \darr\gamma_\mu + a_2\gamma_\mu\darr) H^T L^c
        \partial^\mu\varphi \eeq 
These are the lowest dimension operators that are possible; note that they are
suppressed by two more powers of the heavy neutrino mass $M$ than are the OMM
effective couplings. It follows that $\Sca(\bbcm)$ is suppressed by at least
the factor $\theta^2 q p/M^2$ relative to the corresponding OMM result, where
$q$ and $p$ are respectively the average momenta of the majoron and virtual
neutrino. 

This estimate gives us constraints on the parameters needed if the underlying
heavy-neutrino model is to reproduce the observed anomalous $\bb$ events.
Recall the OMM result, $\geff(\hbox{OMM}) \sim g \theta^2$, where $g$ and
$\theta$ respectively measure the couplings between the heavy neutrino and the
majoron, and its mixing with $\nu_e$.  Roughly, the corresponding CMM
result is
\label\cmmbbdecay
\eq
\geff(\hbox{CMM}) \sim g \theta^2 \; \left( {Q \pf \over M^2} \right). \eeq
Using $Q \sim 1$ MeV and $\pf \sim 100$ MeV, and $\theta < 0.1$, we see that
$\geff \sim 10^{-4}$, as required, only if ($i$) $g \sim 1$, and
($ii$) $\pf/M \sim O(1)$. 

\subsection{A Renormalizable Model}

The above considerations may be simply illustrated within a renormalizable
model. We must first choose the global symmetry group $G$ that will break to
give a majoron carrying a $U(1)$ charge.  The standard model itself provides us
with an example, since if \gwk\ were a global rather than a gauged symmetry,
the \ngb s eaten by $W^+$ and $W^-$ would each carry a unit of electric charge.
We are therefore led to try an analogous global symmetry \gfl, which is to be
broken down to ordinary lepton (electron) number $U_{\sss L}(1)$ by scalar fields
$\phi_i$. These are like the two components of the standard model Higgs in
being a doublet under the new $SU_{\sss F}(2)$ symmetry; however they are gauge
singlets. In further analogy, $\phi_i$ carries a unit of the $U_{\sss L'}$
charge, just as the standard-model Higgs carries weak hypercharge. Our field
content is completed by adding an $SU_{\sss F}(2)$ doublet of  right-handed
gauge-singlet neutrinos, $N_\pm$, and two sterile $SU_{\sss F}(2)$-singlet
neutrinos $s_{\sss\pm}$ carrying only the new $U_{\sss L'}$ quantum number,
namely $L'=\pm 1$.  The $U_{\sss L'}(1)$ factor is required to permit lepton
number to be embedded into the flavour group through the mixing of $\nu_e$ and
the new singlets.  Explicitly, the transformation properties of the new fields
under \gfl\ are  
\label\quantumnumbers
\eq \Pr N \equiv { N_- \choose N_+} \sim ({\bf 2}, 0); \qquad 
       \Pr s_{\sss\pm} \sim ({\bf 1}, \pm 1);  \qquad  \Phi \equiv { \phi_{--}
        \choose  \phi_0} \sim ({\bf 2}, -1), \eeq
where the subscripts denote the corresponding charges under the ultimately
unbroken lepton number, $L = -2 T_3 + L'$.

We construct the most general renormalizable lagrangian respecting all the
symmetries.  The usual standard model particles are taken to be singlets under
$SU_{\sss F}(2)$, and their $U_{\sss L'}(1)$ quantum numbers are chosen to
coincide with their lepton (or electron) number.  The new mass terms and Yukawa
couplings are 
\label\cmmlagrangian
\eq  \Scl =  - \lambda \bar{L}H\Pr\sm  - M \bar\sp \Pr \sm - 
         g_+ \, (\bar{N} \Pl \sp) \; \Phi - g_- \, (\bar{N} \Pl \sm) \; 
        \tw{\Phi} + c.c. \eeq
(A direct mass term for $N$ is forbidden by the $SU_{\sss F}(2)$ symmetry.)
Here $\tw{\Phi} = i \tau_2 \Phi^*$ is the conjugate $SU_{\sss F}(2)$ doublet,
with $\tau_2$ the second Pauli matrix acting on flavour indices. 
The scalar potential is chosen to ensure that $\Phi$ gets a VEV, which we
assume has been rotated to the form
\label\gflvevs
\eq  \Avg{ \Phi} = { 0 \choose u}, \eeq
This breaks \gfl\ down to \ul, with the unbroken electron-type lepton number
symmetry generated by $L$. 

The mass matrix resulting from eqs.~\cmmlagrangian\ and \gflvevs\ yields a
massless neutrino, $\nu_e'$, and two heavy Dirac neutrinos, $\psi_\pm$, whose
masses can be written as 
\label\Diracmasses
\eq \eqalign{
M_\pm &= \hf \left\{ \wtM^2 \pm \sqrt{\wtM^2-4\gp^2u^2\left(\lambda^2 v^2
+\gm^2u^2\right)}\right\}^{1/2} \cr
\hbox{with} \qquad \wtM^2 &= M^2 + \lambda^2v^2 + (\gm^2 + \gp^2) u^2\cr}. \eeq
In terms of left-handed neutrino fields, $\nu_e'$, $\psi_+$ and $\psi_-$ carry
$L=+1$, while $\psi_+'$ and $\psi_-'$ carry $L=-1$. Only the $L=+1$ fields
mix with the electroweak eigenstate, $\nu_e$, with a mixing matrix given by
\label\cmmmixing
\eq \nu_e = \nu_e' c_\theta + (\psi_+ s_\alpha + \psi_- c_\alpha)
s_\theta, \eeq
with $s_\theta = \sin\theta$, $c_\theta = \cos\theta$, \etc\ denoting mixing
angles which are given in terms of model parameters by: 
\eq \eqalign{
\tan\theta &= {\lambda v \over \gm u}, \cr
\tan 2\alpha &= { 2M\sqrt{ \lambda^2 v^2 + \gm^2 u^2 }\over 
M^2 - \lambda^2 v^2 + (\gp^2 - \gm^2) u^2}. \cr} \eeq

We are interested in the couplings $b_{ij}$ controlling the $\bbcm$ decay.  
Strictly speaking this is a $5\times 5$ matrix since there are five left-handed
neutrinos, but because lepton number conservation only permits the $b_{ij}$
coupling among $L=+1$ neutrinos, the result can be simply expressed in terms
of $\nu_e'$ and $\psi_\pm$. In the basis $(\nu_e', \psi_-, \psi_+)$ we have
\label\cmmcouplings
\eq
b_{ij} = {i\gp\over 2}\pmatrix{ 0 & s_\theta s_\alpha & -s_\theta c_\alpha \cr
s_\theta s_\alpha & - c_\theta\sin 2\alpha & c_\theta\cos 2\alpha \cr 
-s_\theta c_\alpha & c_\theta \cos 2\alpha & c_\theta\sin 2\alpha \cr} . \eeq
This expression has the property that the $\bbcm$ amplitude is zero if either
of the couplings $\gp$ or $\gm$ vanishes.  

In addition to the charged majorons, there is also a neutral one, $\varphi_3$,
corresponding to the diagonal generator of $SU_{\sss F}(2)$. Its Yukawa
couplings to neutrinos can be read directly from the lagrangian once this is
expressed in terms of neutrino mass eigenstates. The coupling is
\eq
\Scl_{3\nu\nu} = - { c_{ij} \over 2} \; \ol{\nu}_i \Pl \nu_j \varphi_3 + \cc,
\eeq
where $c_{ij}$ is a 3-by-2 matrix whose rows are labelled by the $L=+1$ states
($\nu_e'$, $\psi_+$, $\psi_-$), and whose columns are labelled by the $L=-1$
states ($\psi_+'$, $\psi_-'$).  We find 
\label\neutralcm
\eq 
c_{ij} = {1\over\sqrt{2}}
\pmatrix{ \gm s_\theta s_\beta & -\gm s_\theta c_\beta \cr
- \gm c_\theta c_\alpha s_\beta - \gp s_\alpha c_\beta & \gm c_\theta
c_\alpha c_\beta - \gp s_\alpha s_\beta \cr - \gm c_\theta s_\alpha s_\beta
+ \gp c_\alpha c_\beta & \gm c_\theta s_\alpha c_\beta + \gp c_\alpha s_\beta
\cr} . \eeq
$\beta$ here denotes the mixing angle amongst the $L=-1$ fields, and is given
in terms of the model parameters by:
\eq 
\tan 2\beta =  { 2\gp uM\over M^2 +\lambda^2 v^2 + (\gm^2-\gp^2) u^2}. \eeq

We should remark that the mass terms in the $\psi_\pm$, $\psi'_\pm$ field 
variables have the form $\hf M_\pm\bar\psi_\pm\psi'_\pm + {\rm h.c.}$, leading 
to matrix propagators 
\label\prop
\eq
\pmatrix{\vev{\psi_\pm\bar\psi_\pm} & \vev{\psi_\pm\bar\psi_\pm'} \cr
	\vev{\psi_\pm'\bar\psi_\pm} & \vev{\psi_\pm'\bar\psi_\pm'} \cr}
= {i\over p^2 - M_\pm^2}\pmatrix{\psl & M \cr M & \psl \cr}. \eeq
The usual Dirac mass terms $M_\pm\bar\psi_\pm\psi_\pm$ and propagators
$\vev{\psi_\pm\bar\psi_\pm} = i/(\psl - M_\pm)$ can be recovered by making the
transformation 
\label\newfields
\eq
\pmatrix{\psi_\pm\cr \psi_\pm'} \to \pmatrix{\Pr & \Pl \cr \Pl & \Pr \cr}
\pmatrix{\psi_\pm\cr \psi_\pm^c} \eeq
where $\psi^c$ denotes the usual charge conjugate field.  In these more 
conventional variables, only the chirality projections of $\psi_\pm$ have 
definite lepton numbers.

\subsection{Naturalness} 

The naturalness issues are much less severe in this model than they are for
OMM's. This is because the unbroken lepton number permits the scale, $u$, of
$SU_{\sss F}(2)$-breaking to be much higher than $O(10\hbox{ keV})$ without
inducing unacceptably large $\bbzn$ decay. The first step is to determine how
large this scale can be. We saw earlier in this section that the conditions for
achieving an acceptable $\bbcm$ rate require heavy neutrino masses, $M_\pm \sim
\pf \sim 100$ MeV, with comparatively large heavy-neutrino scalar couplings,
$g_\pm \sim 1$. We also found that $g u \sim M$ if the neutrino-scalar coupling
is not to be suppressed by additional mixing angles, such as $\alpha$ of our
explicit example. Taken together, these conditions imply an $SU_{\sss F}(2)$
symmetry-breaking scale, $u \sim \pf \sim 100$ MeV. 

Besides being four orders of magnitude larger than the symmetry-breaking scale
that is permitted for OMM's, the CMM scalar sector is also more natural for
another reason. In both cases the largest contribution to the scalar potential
comes from loops which involve the heavy neutrino, of mass $M$. This
contribution is dangerous for OMM's because this neutrino is itself much
heavier than the lepton symmetry-breaking scale. The same is not true of CMM's,
however, because for these models both $u$ and $M$ are of the same size. As a
result, even though these particles couple with nonnegligible strength, $g \sim
1$, the contributions of heavy-neutrino loops are
\eqa
\delta \rho^2 &\sim {g^2 M^2 \over 16\pi^2}, \eolnn
\delta \xi &\sim {g^2 \lambda^2 \over 16\pi^2}. \eeol \eeq
$\delta \rho^2$ is clearly acceptably small, since all that is required is
$\delta\rho^2 \lsim M^2$. The majoron-Higgs coupling, on the other hand, must
satisfy $\delta \xi \lsim 10^{-6}$, which is also easily satisfied given the
phenomenological constraint that $V_{ei} \sim \lambda v /M \sim 0.1$, which
implies $\lambda \sim 10^{-4}$.

\section{Other Bounds}

The couplings of majorons to matter are seen most directly in the $\bb$
processes which have been the main subject of this work, but there are other
constraints which must also be considered.  The most serious of these are
laboratory searches for the mixing of $\nu_e$ with a heavy neutrino, which is
one of the generic predictions of the models we have discussed above.  In
addition, one must take care that majoron emission from stars or supernovae
does not cause them to burn out prematurely, nor do majorons make so large a
contribution to the energy density of the universe that they cause too much
helium synthesis or cause the Hubble expansion to slow too much.  In the
following sections we discuss the models proposed above with regard to
these issues. 

\subsection{Laboratory Bounds}

It was argued that in order to get an observable rate of $\bbom$ or $\bbcm$
events, it is necessary to have a neutrino $\nu_h$ in the 100 MeV range which
mixes with $\nu_e$.  Such a neutrino could be inferred from a `spike' it
implies for the positron spectra of the decays $\pi^+, K^+\to e^+\nu_h$ if it
is lighter than the decaying meson.  A survey of the Particle Data Book 
\ref\PDB{Particle Data Group, \prd{45}{92}{Part 2}.}
\PDB\ shows that pion decay experiments limit the mixing angle to values
$\theta<10^{-3}$ for a neutrino with mass $M_h\sim 100$ MeV. On the other hand,
the above analysis indicates that the minimum angle needed for observable $\bbm$
is approximately $0.1$ for charged majorons and $0.01$ for ordinary majorons.

The bound on the mixing angle from pion decays is easily evaded by taking
$M_h>m_\pi$ so that the decay is kinematically forbidden.  Note that
experimental constraints on mixing coming from searches for the decays of
$\nu_h$ do not apply to our models.  These constraints assume the visible decay
channel $\nu_h\to e^+e^-\nu_e$ due to weak interactions, but in the present
situation the weak process is completely subdominant to decays into majorons,
$\nu_h\to\nu_e\varphi$, which would be undetectable.   

One must therefore look to the decays $K\to e\nu_h$ for limits on the mixing
angle when $M_h>140$ MeV.  The Particle Data Book lists such constraints only
up to a mass of 160 MeV, so one might be misled into thinking that a modest
increase in $M_h$ above $m_\pi$ would render large mixing angles safe from
\ref\KEK{T.~Yamazaki \etal, proceedings of XIth International Conference on
Neutrino Physics and Astrophysics, Dortmund, West Germany, Jun 11-16, 1984}
being ruled out. Actually there exist stringent results from KEK \KEK\ that do
not appear in ref.~\PDB. This experiment also restricts $\theta\lsim 10^{-3}$
for masses up to $M_h = 350$ MeV.  Such a large mass leads to a large
suppression of the amplitude for $\bbcm$, although not necessarily for $\bbom$.

An indirect limit on the coupling of $\nu_e$ to heavy neutrinos also comes from
tests for universality of the weak interactions of leptons in different 
families.  The most restrictive test comes from the comparison of electron and 
muon charged current couplings in pion decays 
\ref\Bryman{D.A.~Bryman, \prd{35}{92}{1064}.}
\Bryman. Suppose that $\nu_e$ had mixing angle $\theta$ to a neutrino with mass
$M > m_\pi$, so that its weak couplings were suppressed  relative to those of
$\nu_\mu$ by $\cos\theta$ (assuming for simplicity that $\nu_\mu$ does not mix
with anything.)  The comparison of theory with experiment shows that the ratio
of electron to muon couplings measured in meson decays is 
\label\universality
\eq { (\GF)_e \over (\GF)_\mu} = \cos\theta = 0.9970\pm 0.0023.\eeq
Taking the one-sigma lower deviation we get a bound on the mixing angle of
\label\unibound
\eq \theta < 0.10, \eeq
which is marginally consistent with having observable $\bbm$ in our models. 
Note that a real deviation of \universality\ from unity would be indirect
evidence for the sort of neutrino mixing we need. 

A further constraint on the majoron coupling to neutrinos comes from searches
for the decay $\pi\to e\nu\varphi$ 
\ref\Pic{C.E.~Picciotto \etal, \prd{37}{88}{1131}.}
\Pic. These yield a comparatively weak limit of $\geff < 9\times 10^{-3}$. 

\subsection{Cosmology and Astrophysics}

Because of the weak coupling of the majoron to matter, one might worry that it
could have deleterious cosmological effects, such as contributing too much
energy density if it is massive, interfering with the formation of large scale
structure due to its decays.  Emission of majorons from stars or supernovae
might also shorten the lifetimes of either.

In fact the effective coupling $\geff\sim 10^{-4}$ of a massive majoron to
neutrinos is sufficient for avoiding the cosmological problems. The majoron
lifetime due to the decay $\varphi \to \nu_e\nu_e$ is 
\label\majoronlifetime
\eq \tau_\varphi = 16\pi\geff^{-2} m_\varphi^{-1}, \eeq
which is $10^{-10}$ s for $m_\varphi\sim 10$ keV, far less than is required by
consideration of the density of the universe or galaxy formation.

In contrast, the majoron coupling to ordinary matter such as found in stars is
too {\it weak} to do any harm.  Since the thermal background of neutrinos in a
star is negligible, majorons are emitted primarily as Bremsstrahlung from
electrons.  But for the OMM considered above, lepton number conservation
prevents a single majoron from being emitted; rather they must appear in pairs
with zero net lepton number.  The effective coupling of two majorons to
electrons is generated by a loop diagram in which a $W$ boson is exchanged,
Fig.~(6).  The resulting effective interaction with electrons can be estimated
as
\label\twophi
\eq {\theta^2 m_e \GF\over 16\pi^2} |\varphi|^2 \bar e\gamma_5 e,
\eeq
where $\theta$ is the mixing angle between $\nu_e$ and the heavy neutrino,
whose mass does not appear because we have assumed it to be much less than
$\GF^{-1/2}$.  The amplitude for $\varphi$ emission proves to be some eight
orders of magnitude below the observational limit.   We expect similar results
for charged majorons, which must also be emitted in pairs.  But in addition we
need to check the rate of neutral majoron ($\varphi_3$) emission in the CMM. 
The coupling of $\varphi_3$ to electrons arises at one loop from $W$ and $Z$
boson exchange.  To make an estimate we have computed only the latter
contribution (the two are numerically equal in the singlet majoron model \CMP).
It is shown in Fig.~(7).  Using the couplings of eq.~\neutralcm, one can
eventually find that the effective interaction has the form
\label\onephi
\eq     {\lambda^2 m_e \over 16\sqrt{2} \,\pi^2 u} f(\theta,\alpha,\beta)
        \left[(1+2\epsilon)\ln(1+\epsilon) - 2\right]\;
	 \varphi \; \bar e\gamma_5 e,
        \quad \epsilon = (\Mp^2-\Mm^2)/\Mm^2 \eeq
where $f(\theta,\alpha,\beta)$ is a function of the three mixing angles of the
model --- see Section (5.1) for their definitions --- and which we here
conservatively take to be $O(1)$. Recall that $\lambda v \sim \theta_i  M_i\sim
10$ MeV from the requirement of getting observable $\bbcm$.  Using the fact
that $M_i\sim u$, we get a coefficient of order $10^{-13}$.  Comparing with the
analysis of ref.~\georgi, one sees that this is somewhat below the limit from
red giant lifetimes of $10^{-6}$ times the electron Yukawa coupling, or
$3\times 10^{-12}$. 

Because of the higher temperatures in supernovae, weak interactions are in
equilibrium and there is a thermal population of neutrinos.  A coupling of
order $\geff=10^{-4}$ between neutrinos and majorons is sufficient for bringing
the latter into equilibrium as well 
\ref\Choi{K.~Choi and A.~Santamaria, \prd{42}{90}{293}.} 
\Choi. Therefore, in contrast to the situation for stars, in supernovae majorons
are so strongly coupled that they are {\it trapped} in the core, and do not
significantly deplete the normal energy flux, in this case due to neutrinos. 
This will be made more quantitative in the next section, where we examine the
equilibration of majorons when the universe was at a temperature of $1-100$ MeV:
conditions similar to those in a supernova. 

If majorons are trapped in supernovae they can have an adverse effect on the
bounce and subsequent explosion 
\ref\Fuller{G.M.~Fuller, R.~Mayle and J.R.~Wilson, Astrophys.~J.~{\bf 332}
(1988) 826.}
\Fuller. This has only been studied for triplet majorons, using restrictive
assumptions about the energy dependence of the cross section for
$\nu\nu\to\varphi\varphi$, so that no direct conclusions on the models studied
here can be drawn.

\subsection{Nucleosynthesis}

A difficulty not so easily surmounted is that the majorons in our models
generally change the expansion rate of the universe enough to have increased
the predicted abundance of primordial Helium 
\ref\Chang{S.~Chang and K.~Choi, Seoul preprint SNUTP 92-87.}
\Chang.  We will show how this comes about and suggest some possibilities for
evading the problem. 

Every scalar degree of freedom in equilibrium at MeV temperatures in the early
universe is equivalent to $4/7$ of a neutrino species in its contribution to
the energy density, and hence the expansion rate.  A complex scalar, as in the
OMM we have discussed, would thus count as $8/7$, and the CMM would give $12/7$
because  it has a total of three \ngb s.  The current limit on the number of
additional neutrino species beyond those of the standard model is $0.4$
\ref\WSSOK{ T.P.~Walker, G.~Steigman, D.N.~Schramm, K.A.~Olive and H.S.~Kang,
Astrophys. J. {\bf 376} (1991) 51.}
\WSSOK. 

In the OMM's, the dominant means for equilibrating massive majorons is the
decay $\varphi\to\nu\nu$ and its inverse process.  The thermally averaged rate
is roughly 
\label\invdecay
\eq \Gamma \sim 10^{-2} \geff^2 m_\varphi^2 /T, \eeq
which comes into equilibrium before a temperature of 1 MeV for all scalar
masses greater than a few eV, assuming $\geff=10^{-4}$.  Since it becomes
increasingly unnatural to have scalars lighter than the 10 keV allowed by the
$\bb$ experimental anomaly, we expect the decays to be in equilibrium for
massive majorons. 

Charged majorons will suffer fast decays only if they develop a large enough
thermal mass.  Rather than compute this, we focus on the annihilation process 
$\nu\bar\nu\to\varphi\varphi^*$.  Using the interactions of eq.~\cmmcouplings,
we estimate the thermally averaged annihilation rate to be of order
\label\nunutocm
\eq \Gamma \sim {1\over\pi^3}\theta^4 T^5 M^{-4} \eeq
in the limit that $T\ll M$.  This is some ten orders of magnitude faster than
the expansion rate at temperatures of an MeV, assuming masses and mixing angles
of $\theta\sim 0.1$ and $M\sim 100$ MeV.  In addition, using the neutral
majoron interactions of eq.~\neutralcm, we find that the rate for
$\nu\nu\to\varphi\varphi_3$ can be suppressed relative to \nunutocm\ only by a
factor of $(T/M)^2$.  Thus all three kinds of majorons will be in equilibrium
at $T\sim 1$ MeV, in contradiction to the nucleosynthesis bound.  

We would like to point out two ways in which the nucleosynthesis may proceed as 
usual, despite the presence of two or three majoron species.  One possibility
is that the tau neutrino mass is close to its experimental upper bound, in the
region of 5 to 30 MeV.  If it decays or annihilates into majorons on time 
scales faster than 1 s, the time when neutrinos decouple, there will be
one less species of neutrinos, making room for two species of majorons, or
three with a weak violation of the bound.  

A second possibility is that some neutrino decays into $\nu_e$ plus $\varphi$
in such a way as to heat the electron neutrinos relative to the other species.
It was shown that this occurs if $\nu_\mu$ or $\nu_\tau$ has the desired decay 
with a lifetime in the range $6\times10^{-4}$ s $< \tau < 2\times 10^{-2}$ s
\ref\kimmo{K.~Enqvist, K.~Kainulainen and M.~Thomson, \prl{68}{92}{744}.}
\kimmo.  The overpopulation of $\nu_e$  results in prolonged equilibrium
between neutrons and protons, which compensates for the extra density of the
universe in its effect on helium synthesis.  This idea can be generalized to
the annihilations of sterile neutrinos in the mass range of a few MeV as well.
In fact it is not necessary that the decaying or annihilating particle go 
directly into $\nu_e$'s; as long as it produces particles that are in 
equilibrium with $\nu_e$, after the decoupling of neutrinos from electrons, it 
will accomplish the same thing.

As an existence proof for these mechanisms, we show how the ordinary majoron
model of section (4.2) can be generalized to include a heavy tau neutrino.
Let there be one additional sterile neutrino, $s_3$, whose lepton number is the
same as that of $\sm$.  When we include the other two generations, the
straightforward extension of the lagrangian \ommrenmodel\ yields a mass matrix
of the form
\label\bigmatrix
\eq \pmatrix{ \spa & \spa & \spa & m_1 & 0 & m_4 \cr 
              \spa & 0    & \spa & m_2 & 0 & m_5 \cr 
              \spa & \spa & \spa & m_3 & 0 & m_6 \cr
              m_1 & m_2 & m_3 & 0 & M_1 & 0 \cr  
              0 & 0 & 0 & M_1 & 0 & M_2 \cr  
              m_4 & m_5 & m_6 & 0 & M_2 & 0 \cr} \eeq
in the basis $(\nu_e,\nu_\mu,\nu_\tau,\sm,\sp,s_3)$.  It is easy to see that
the spectrum consists of two massless states which are mostly $\nu_e$ and
$\nu_\mu$, a Dirac neutrino of mass $\sim m_i$ consisting mostly of $\nu_\tau$
and $\sm$, and a Dirac neutrino of mass $\sim M_i$ which is mostly $\sp$ and
$s_3$.  Supposing that the intermediate Dirac mass is of order 10 MeV, for 
example, we see that the constraints on mixing angles can be satisfied:
\label\etaumix
\eq \theta_{e\tau} = (cm_1 - sm_4)/(cm_3-sm_6) \ltwid 0.01 \eeq
from searches for peaks in the $\pi\to e\nu$ spectrum \Bryman, and 
\label\esmix
\eq \theta_{es} = \left[(cm_4+sm_1) -\theta_{e\tau}(cm_6+sm_3)\right/M \sim
0.01 \eeq
from the requirement that $\geff\sim 10^{-4}$ in $\bbom$.  Here $s/c = M_1/M_2$
is the tangent of the mixing angle in the sterile neutrino sector,
$M=(M_1^2+M_2^2)^{1/2}$ is approximately the mass of the heaviest state, and
$\theta_{ex}$ denotes the mixing angle between $\nu_e$
and the mass eigenstates that are mostly $\nu_\tau$ or the heavy sterile
neutrino. 

It turns out that the tree level couplings that would cause the decay
$\nu_\tau\to\nu_e\varphi$ vanish; nevertheless the annihilation process 
$\nu_\tau\bar\nu_\tau\to\varphi\varphi^*$ goes at a rate comparable to 
\nunutocm, however without the mixing angle suppression.  The annihilations are 
therefore very efficient in depleting the $\nu_\tau$ population, as long as the 
heavy neutrino mass scale $M$ is significantly smaller than 100 GeV.  Moreover
the resulting majorons are still in equilibrium with the light neutrinos, so we
have the $\nu_e$-heating mechanism in addition to the elimination of
$\nu_\tau$.  We note that a tau neutrino in this mass range would not 
necessarily have manifested itself in supernova 1987a through the delayed 
signal of its decay products.  Because it interacts so strongly with majorons, 
its neutrinosphere will be farther out in the core where the temperature is 
lower and the Boltzmann suppression is greater, contrary to the usual case 
where $\nu_\tau$ is emitted at a higher temperature.  Thus the flux of 
$\nu_\tau$'s would be greatly reduced relative to the electron neutrinos.

\section{Conclusions}

Motivated by experiments suggestive of majoron emission in double beta 
decay, we have proposed two kinds of models that are able to account for this 
effect without lepton number violation.  In the first proposal, the boson is
not of the Nambu-Goldstone variety but rather has a small mass, which can
nevertheless be natural in a technical sense discussed above.  The second
proposal is to let the majoron be truly massless, but carry lepton number 
charge.  Coincidentally, both of these schemes suggest the existence of heavy
isosinglet neutrinos in the mass range of several hundred MeV with significant
mixing to the electron neutrino.  These heavy neutrinos could manifest
themselves in the decays $K\to\nu e$ or by nonuniversality in the weak
interactions of electrons versus other leptons.  The models can be consistent 
with nucleosynthesis constraints if the tau neutrino is in the $1-10$ MeV 
mass range, or there exist additional sterile neutrinos with a mass of a few 
MeV.  The anomaly in the double beta decay spectra, if confirmed, would thus be 
the precursor to several new phenomena in neutrino physics.
 
\bigskip
\centerline{\bf Acknowledgments}
\bigskip

This research was partially funded by funds from the N.S.E.R.C.\ of Canada,
les Fonds F.C.A.R.\ du Qu\'ebec, and DOE grant DE-AC02-83ER-40105.

\bigskip
\centerline{\bf Note Added}
\bigskip
Since completing this work we have been informed of evidence that the anomalous 
events reported by the UC Irvine group may be due to resolution problems for 
the higher-energy electrons \ref\MoeII{M.~Moe, private communication of results 
to appear in Int.~J.~Mod.~Phys.~{\bf E} (1993).}\MoeII.

\appendix{A}{The Yukawa formulation of NGB Couplings}

In this Appendix we derive the general form for the Yukawa coupling of a \ngb:
eq.~\ngbyukawas. 

To this end consider an arbitrary set of Yukawa couplings between a collection
of spin $1/2$ and spin $0$ particles. Such particles may always be cast as
Majorana fermions, $\psi^i$, and real scalar fields, $\Phi_a$. The
most general form for their mutual couplings is
\label\generalcouplings
\eq  \Scl_{\rm yuk} = - \hf \; \bar{\psi}^i \Gamma^a_{ij}\, \Pl  \psi^j \;
\Phi_a
        + c.c.  \eeq

Suppose also that this lagrangian is invariant with respect to the following
global symmetry transformations:
\label\symmetry
\eqa \delta \psi^i  &= i \theta^\alpha \left[ {(q_\alpha)^i}_j \Pl  - 
        {{(q_\alpha)^i}_j }^*  \Pr \right] \, \psi^j  \eolnn  
        \delta \Phi_a &= i \theta^\alpha {(\Scq_\alpha)_a}^b \Phi_b, \eeol \eeq
in which both sets of matrices, $q_\alpha$ and $\Scq_\alpha$, are hermitian, 
and $\Scq_\alpha$ must also be imaginary.  Invariance of the Yukawa couplings
is expressed by the identity
\label\identity
\eq    (q_\alpha)^\transp \, \Gamma^a +  \Gamma^a  \, q_\alpha  +
        \Gamma^b \, {(\Scq_\alpha)_b}^a \equiv 0.  \eeq
Any explicit left-handed fermion mass matrix, $(m_0)_{ij}$,  must similarly
satisfy the relation $q^\transp m_0 + m_0 q \equiv 0$.

This symmetry is spontaneously broken when the scalar fields acquire their
VEV's, $v_a = \Avg{\Phi_a}$, and the resulting \ngb\ directions in scalar-field
space, $\varphi_\alpha$, are given by the action of the symmetry on $v_a$: 
\label\gbdirection
\eq  (\delta_{\sss GB} \Phi)_a  \equiv  i   {( \Scq_\alpha v)_a}  \;
        (F^{-1})^{\alpha\beta} \; \varphi_\beta. \eeq
The real, symmetric normalization matrix, $F^{-1}$, is chosen to 
ensure that the scalar kinetic terms remain properly normalized. 
That is, $\partial_\mu \Phi_a \, \partial^\mu \Phi_a = \partial_\mu
\varphi_\alpha  \, \partial^\mu \varphi_\alpha + \cdots$, provided that
\label\decayconstant
\eq  (F^{-1})^{\alpha\gamma} \; \left[ v^\transp \Scq_\gamma 
        \Scq_\lambda v \right] \; (F^{-1})^{\lambda\beta} =
\delta^{\alpha\beta}. \eeq

The Yukawa coupling for $\varphi_\alpha$ therefore becomes
\label\yukawaresult
\eqa    \Scl_{\rm yuk}& = -  {i\over 2}  \; \bar{\psi}  \, \Gamma^a \, \Pl  \psi
        \; (\Scq_\alpha v)_a \;  (F^{-1})^{\alpha\beta} \; \varphi_\beta + c.c.
         \eolnn
        &= + {i\over 2}  \; \bar{\psi} \left(  q_\alpha^\transp \Gamma^a + 
        \Gamma^a  q_\alpha \right) \Pl  \psi \;  v_a \; (F^{-1})^{\alpha\beta} 
        \; \varphi_\beta + c.c.,  \eeol  \eeq
where eq.~\identity\ was used in writing the last line. The expression
for the right-handed coupling follows simply from taking the complex 
conjugate of this expression.

Eq.~\yukawaresult\ gives the most general form for \ngb\ couplings.  It can be
recast into the form of eq.~\ngbyukawas\ using some additional simplifying
features of the models which we consider. Suppose first that  no 
symmetry-invariant fermion mass terms exist, $m_0 = 0$. Then the Yukawa 
coupling matrices, $\Gamma^a$, which appear in eq.~\yukawaresult\ can be 
traded for a dependence on the fermion mass matrix using
\label\mintermsofgamma
\eq  m_{ij} = \Gamma^a_{ij} \; v_a. \eeq

Next, suppose that the \ngb s carry an unbroken $U(1)$ charge, as is the case
for CMM's. It is then convenient to work with complex combinations of the
$\varphi_\alpha$'s.  If, for example $\varphi_1$ and $\varphi_2$ form a
multiplet under the unbroken $U(1)$, then the symmetry transformations become
diagonal when expressed in terms of $\varphi = (\varphi_1 +  i
\varphi_2)/\sqrt{2}$.  The same steps as before once more lead to
eqs.~\yukawaresult, with the proviso  that the corresponding broken charge, $q
= (q_1 - i q_2)/\sqrt{2}$,  need no longer be hermitian. 

The simplest case is if there is only one \ngb\ with a nonzero charge, as in
the models we consider. Then the normalization matrix, $F^{-1}$, cannot mix
$\varphi$ with any of the uncharged \ngb s, and must be proportional to the
unit matrix in the charged-scalar sector. Denoting the proportionality constant
by: $1/f  = (F^{-1})^{11} =  (F^{-1})^{22}$, we obtain eqs.~\ngbyukawas, as
required. 

\appendix{B}{The Equivalence of Derivative and Yukawa Formulations}

A famous property of \ngb s is that they only couple derivatively. Here we make
this property explicit for the couplings of the \ngb s that are considered in
eq.~\ngbyukawas\ by showing the equivalence of these two formulations for the
double-beta decay rate.

Nambu-Goldstone bosons can only couple derivatively because if these fields are
taken to be constants they completely drop out of the lagrangian density. This
is because the Nambu-Goldstone directions in field space are defined by
performing a field-dependent symmetry tranformations on the vacuum, as in 
eq.~\gbdirection.  For constant fields these transformations are really
symmetries, and so produce no effect at all in the lagrangian. $\varphi_\alpha$
only appears to the extent that it varies in spacetime, and so it must couple
only through its derivatives.

To see this in the present case, consider the following field-dependent
redefinition of the fermion fields: 
\label\redefinition
\eq  \delta \psi  \equiv  -i  (q_\alpha \Pl - q^\transp_\alpha \Pr) \psi  \;
        (F^{-1})^{\alpha\beta}  \; \varphi_\beta. \eeq
The fermion mass term changes by
\label\deltamass
\eq  \delta \Scl_{\rm mass} =  -i \bar{\psi} ( q_\alpha^\transp m + m q_\alpha)
        \Pl \psi \;  (F^{-1})^{\alpha\beta} \; \varphi_\beta + c.c., \eeq
where $m = \Gamma^a v_a$. Notice that this is exactly what is required
to cancel the Yukawa coupling of eq.~\yukawaresult.  A similar 
cancellation occurs for {all} of the nonderivative interactions of 
$\varphi_\alpha$. It is important to notice in this regard that if the broken 
symmetry should transform other particles like the electron, in addition to 
neutrinos, then these other particles must also participate in the field
redefinition, eq.~\redefinition, in order to remove all nonderivative 
$\varphi$-dependence. 

The $\varphi_\alpha$-dependence is not completely eliminated, however, since
the fermion kinetic terms are not invariant under a spacetime-dependent
transformation such as that of eq.~\redefinition. It is a simple exercise to
show that, under the assumptions leading to eq.~\ngbyukawas, the variation
of the kinetic term is given by eq.~\derivative. The latter has been
expressed in a way that holds even if the generators, $q$, are not hermitian,
as is appropriate for charged \ngb s. Since these two forms for the \ngb\
interaction are related by a field redefinition, they must give equivalent
scattering amplitudes. 

To concretely verify the equivalence of these two expressions for the
\ngb\ couplings, we compute the double-beta decay rate using both 
eqs.~\ngbyukawas\ and \derivative. Although the graph in which the \ngb\
is emitted by the neutrino line, as in Fig.~(1), is {\it not} equivalent by
itself, we will show that the result becomes equivalent once Fig.~(1) is added
to the remaining graphs of  Fig.~(8). We denote the leptonic part of the 
amplitude computed from Figs.~(1) and (8) using derivative couplings by
$M^{\mu\nu}_\partial$, and the same amplitude using Yukawa couplings by
$M^{\mu\nu}_y$. 

First consider the evaluation of these graphs using derivative couplings. For
generality we work with an arbitrary set of majorona fermions, and real
scalar fields, and assume the following form for scalar and charged-current
interactions which appear in the Feynman rules for Figs.~(1) and (8):
\label\interactions
\eqa  \Scl_{\psi\psi\varphi} &= \bar{\psi}^i \gamma^\mu (\Scv_{ij} + \Sca_{ij}
\,
        \gamma_5) \psi^j \; \partial_\mu \varphi, \eolnn
        \Scl_{\psi\psi W} &= \bar{\psi}^i \gamma^\mu (V_{ij} + A_{ij} \,
        \gamma_5) \psi^j \; W_\mu + c.c., \eeol  \eeq
where, on general grounds, $A$ and $\Sca$ are both symmetric matrices, while
$V$ and $\Scv$ are antisymmetric. For the charged-current weak interactions
we take $A = V \propto T_-$, where $T_-$ is the
$SU_{\sss L}(2)$ lowering operator. For the \ngb\ associated with the charge
$q$, we have $\Sca = \hf (q + q^\transp)$ and $\Scv = \hf (q - q^\transp)$.
The corresponding Yukawa couplings will be denoted by the matrix $\Lambda
\propto a \Pl + b \Pr$, with $a$ and $b$ given by eq.~\ngbyukawas. 
 
Up to a common overall normalization the leptonic part of the integrands for
the three graphs become (in an obvious matrix notation): 
\label\derivintegrands
\eqa  M^{\mu\nu}_\partial(1) &= \bar{u}(p_1) \gamma^\mu 
        (V+iA\gamma_5) S(p_1 - k_1) \qxpsl (\Scv+i\Sca\gamma_5) S(k_2-p_2)
\gamma^\nu
        (V+iA\gamma_5) u^c(p_2), \eolnn
        M^{\mu\nu}_\partial(8a) &= \bar{u}(p_1) \qxpsl 
        (\Scv+i\Sca\gamma_5) S(p_1 +q) \gamma^\mu (V+iA\gamma_5) S(k_2-p_2)
\gamma^\nu
        (V^*+iA^*\gamma_5) u^c(p_2), \eolnn
        M^{\mu\nu}_\partial(8b) &= \bar{u}(p_1) \gamma^\mu 
        (V+iA\gamma_5) S(p_1 - k_1) \gamma^\nu (V^*+iA^*\gamma_5) S(-p_2-q)
\qxpsl
        (\Scv+i\Sca\gamma_5) u^c(p_2), \eolnn & \eeol \eeq
where $S(p) = \left[ i\pxpsl + m\Pl + m^* \Pr \right]^{-1}$ is the fermion
propagator, thought of as a matrix in Dirac and flavour space, while $u$ 
($u^c$) is the (conjugate) electron spinor. 

These expressions can be related to the Yukawa expressions by applying the
following easily proven identities: 
\label\algebra
\eqa S(p_1-k_1) i\qxpsl (\Scv+i\Sca\gamma_5) S(k_2-p_2) &= S(p_1-k_1) \Lambda
        (\Scv+i\Sca\gamma_5) S(k_2-p_2) \eolnn
        & \qquad + S(p_1-k_1) (\Scv+i\Sca\gamma_5) - (\Scv+i\Sca\gamma_5)
        S(k_2-p_2), \eolnn
        \bar{u}(p_1) i \qxpsl (\Scv+i\Sca\gamma_5) S(p_1+q) &= \bar{u}(p_1)
\left[
        (\Scv-i\Sca\gamma_5) + \Lambda S(p_1+q) \right] \eol
        S(-p_2-q) i\qxpsl (\Scv+i\Sca\gamma_5) u^c(p_2) &= \left[
-(\Scv+i\Sca\gamma_5)
        + S(-p_2-q) \Lambda \right] u^c(p_2). \eeolnn \eeq
The last two of these identities rely on using the Dirac equation
for the initial and final spinors, $u(p_1)$ and $u^c(p_2)$. 

Using these identities in eqs.~\derivintegrands\ relates the
derivative-coupling and Yukawa-coupling results for each graph:
\label\comparison
\eqa  M^{\mu\nu}_\partial(1) &= M^{\mu\nu}_y(1) + \bar{u}(p_1) \gamma^\mu
        (V+iA\gamma_5) S(p_1 - k_1)
        (\Scv+i\Sca\gamma_5) \gamma^\nu (V^*+iA^*\gamma_5) u^c(p_2) \eolnn
        & \qquad - \bar{u}(p_1) \gamma^\mu (V+iA\gamma_5) (\Scv+i\Sca\gamma_5)
        S(k_2-p_2) \gamma^\nu (V^*+iA^*\gamma_5) u^c(p_2), \eolnn
        M^{\mu\nu}_\partial(8a) &= M^{\mu\nu}_y(8a) + \bar{u}(p_1) 
	(\Scv-i\Sca\gamma_5)
        \gamma^\mu (V+iA\gamma_5) S(k_2-p_2) \gamma^\nu (V^*+iA^*\gamma_5)
u^c(p_2),
        \eolnn
        M^{\mu\nu}_\partial(8b) &= M^{\mu\nu}_y(8b) + \bar{u}(p_1) \gamma^\mu
(V+iA\gamma_5)
        S(p_1 - k_1) \gamma^\nu (V^*+iA^*\gamma_5) (-\Scv+i\Sca\gamma_5)
u^c(p_2).
        \eolnn & \eeol  \eeq

We see that although the result using the two formulations of the \ngb\
couplings do  {\it not} agree graph by graph, their sum is the same provided
that the scalar- and charged-current coupling matrices satisfy the following
conditions: 
\label\cancellationconditions
\eq [\Scv,V^*] + [\Sca,A^*] = [V^*,\Sca] + [\Scv,A^*] = [\Scv,V] + [A,\Sca] =
        [\Scv,A] + [\Sca,V] = 0.\eeq
These are trivially satisfied if the charged-current generators commute with
the charge that is associated with the \ngb, as is required by the invariance
of the charged-current interactions under the spontaneously-broken global
symmetry. The equivalence of the two formulations for double-beta decay is
thus established. 

\appendix{C}{The C.M. Electron Spectrum}

It is argued in the text that the vanishing of the double-beta decay matrix
element $\Sca(\bbcm)$ as the majoron momentum goes to zero is a key feature
of charged majoron models. Here we demonstrate this property in some detail. 

The vanishing of the amplitude is straightforward when the majoron couplings
are expressed in derivative form, as in eq.~\derivative. In this case the
conservation of electric charge and lepton number precludes any derivative
coupling between the electron and the charged majoron, in the absence of exotic,
electrically-charged fermions. As a result, neither of the graphs of Fig.~(8)
contribute to the CMM double beta decay rate. The remaining graph, Fig.~(1),
manifestly vanishes for zero majoron momentum becaus of the derivative 
coupling.

It is more complicated to see this result in the Yukawa-coupling language.
Lepton number conservation forbids a direct coupling between the charged
majoron and the electron, so the only graph to be considered is again that of
Fig.~(1). As might be expected from Appendix B, however, the result for this
graph need not vanish for zero majoron momenta until the contributions from all
of the relevant intermediate neutrinos have been summed. 

At zero majoron energy, the $\bbm$ decay amplitude is given by 
eq.~\bbomrate, whose integrand is proportional to the $\nu_e$-$\nu_e$
element of the following matrix in flavor space: 
\label\bbomintegrand
\eqa  \hbox{Integrand} & \propto \left[ {1\over p^2 - m^* m} \;\left( m^*
        a m^* + p^2 b \right) \; {1\over p^2 - m m^*} \right]_{\nu_e \nu_e}
\eolnn
        &=  {b_{\nu_e \nu_e} \over p^2} - \sum_{n=0}^\infty { 1 \over
        (p^2)^{n+2}} \left[ (m^* m)^{n+1} \, b + \!\phantom{\sum} \right.\eol
        & \qquad\qquad  \left. \sum_{k=0}^n (m^* m)^k \left(  b m m^* - m^* 
        a m^* \right) (m m^*)^{n-k} \right]_{\nu_e \nu_e}. \eeolnn  \eeq
As in previous expressions, $m = m^\transp$ denotes the complex left-handed
neutrino mass matrix, while $a$ and $b$ are the Yukawa coupling matrices of
eq.~\yukawa.

The last expression simplifies drastically once eq.~\ngbyukawas\ is used, which
contains the information that the majoron is a \ngb. After a pairwise
cancellation of all but one of the terms in the sum over $k$, we find that 
\label\simpleform
\eq  \hbox{Integrand} \propto \left[ {q m^* + m^* q^\transp \over p^2}
        - \sum_{n=0}^\infty { 1 \over (p^2)^{n+2}} \left[ q (m^* m)^{n+1}m^*
        + (m^* m)^{n+1} m^* q^\transp \right] \right]_{\nu_e \nu_e}. \eeq
The significance of this final result lies in the fact that each term in it is
proportional to a $\nu_e$ matrix element, $q_{\nu_e j}$, of the \ngb\ charge.
The final point to be established is that, for CMM's, all such matrix elements
are zero. We are therefore forced to work to next order in the majoron
momentum, eq.~\bbcmamp, in order to get a nonvanishing contribution. 

In order to see why $q_{\nu_e j}$ must vanish in CMM's, consider the
symmetry transformations in the basis of weak-interaction eigenstates. 
Then the invariance of the gauge interactions under the global symmetry
implies that $q$ can only transform the entire doublet, ${\nu_e
\choose e}_{\sss L}$, into other doublets having the same hypercharge. But
since the \ngb\ charge, $q$, has embedded in it two units of the unbroken
lepton number, such a transformation cannot be made without introducing exotic
isodoublet fermions. 

\appendix{D}{The O.M. Electron Spectrum}

Here we wish to show that the OMM double beta decay amplitude, unlike that for 
CMM's, is nonvanishing even at zero majoron momentum. 
This is particularly easy to see using the Yukawa form for the \ngb\
couplings, which can be directly read off from the lagrangian
of a given model.  In this form the majoron typically couples only to
neutrinos, and not to electrons. Thus only the graph of Fig.~(1)
contributes. In contrast to CMM's, it is possible to have $q_{\nu_e j} \ne 0$
in OMM's (see the previous Appendix)  and so the decay rate at zero majoron
momentum need not vanish. 

The puzzle is to understand this result when the amplitude is expressed in
terms of the derivative couplings, since in this formulation all of the graphs
of Fig.~(1) and Figs.~(8) are explicitly proportional to the majoron momentum,
$k$. The resolution turns out to come from the contributions of Figs.~(8). For
these graphs, in which the majoron is emitted from the electron
lines, the internal electron goes on shell in the limit as $k\to 0$, causing a
singularity in the propagator. The coefficient of this singularity is
proportional to the {\it vector} part of the electron-majoron coupling. (The 
same singularity leads to the familiar infrared divergence of the analogous
photon bremsstrahlung graphs in Quantum Electrodynamics.)  Consequently the
electron propagator behaves as $1/k$ for small $k$, which cancels the explicit
$k$-dependence due to the majoron's derivative coupling. 

\listrefs

\vfill\eject
\centerline{\bf Figure Captions}
\bigskip
  
\topic{Figure (1)}
The Feynman graph which gives rise to double-beta decay accompanied by the
emission of a Majoron. The four-fermion vertices are those of the usual
charged-current weak interactions. 

\topic{Figure (2)}
The Feynman graph which gives rise to ordinary two-neutrino double-beta 
decay as occurs in the Standard Model. 

\topic{Figure (3)}
The Feynman graph which gives rise to neutrinoless double-beta decay, with no
Majoron emission. 

\topic{Figure (4)}
The number of decay electrons as a function of the sum of the two electrons'  
energy. The solid curve represents the two-neutrino decay, the dotted curve
gives the OMM decay, and the dashed curve gives the CMM decay.  All three
curves have been arbitrarily assigned the same maximum value for the purposes
of comparison. 
  
\topic{Figure (5)}
The two most dangerous Feynman graphs contributing to the light scalar  
couplings when the heavy neutrino is integrated out. 

\topic{Figure (6)}
The Feynman graph through which the effective electron-Majoron interaction is
induced. 
  
\topic{Figure (7)}
The Feynman graph which mixes the Z boson with the `neutral' Goldstone boson.
Once the Z is attached to a fermion line this induces an effective
electron-Goldstone boson interaction. 

\topic{Figure (8)}
The remaining Feynman graphs which contribute to double-beta decay accompanied
by Majoron emission. These graphs only arise if direct electron-Majoron
couplings exist, as is the case for OMM's in the variables for which the
Majoron is derivatively coupled. 

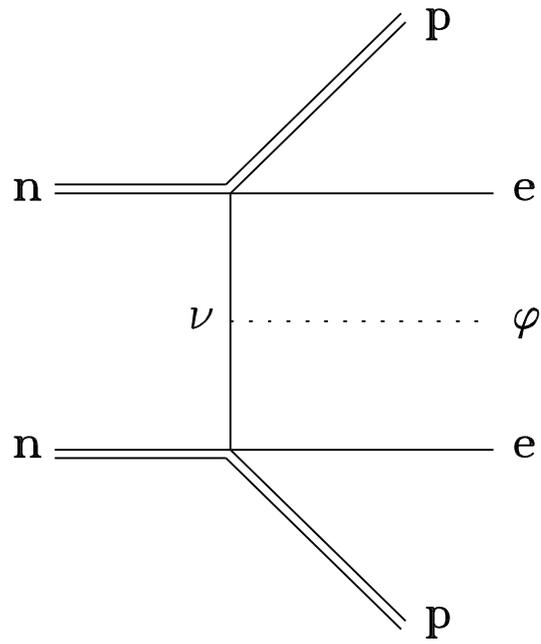
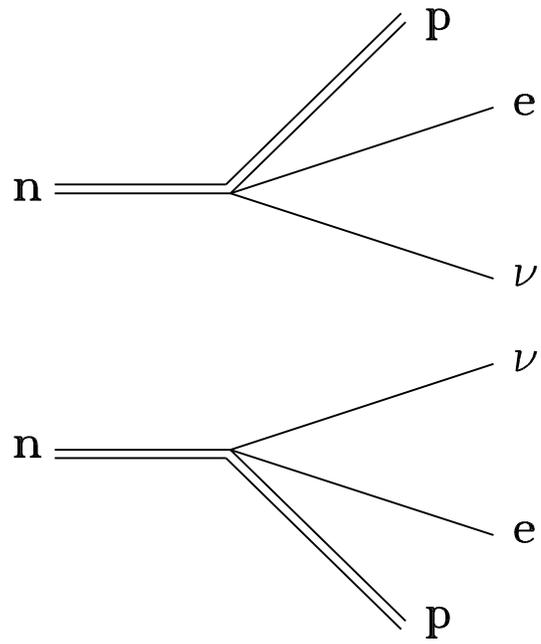
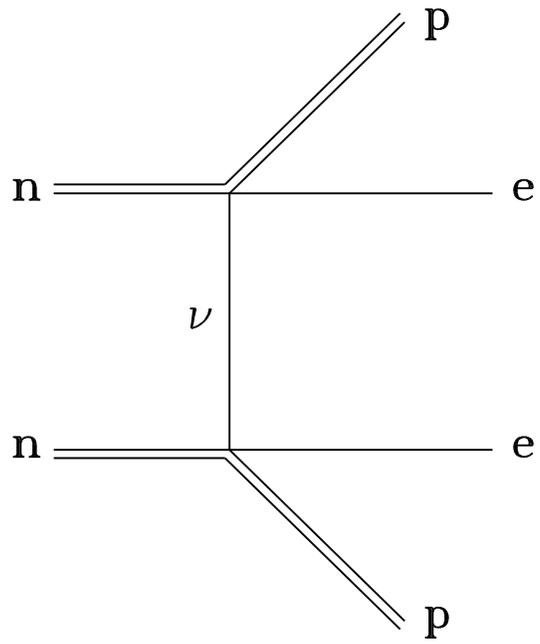
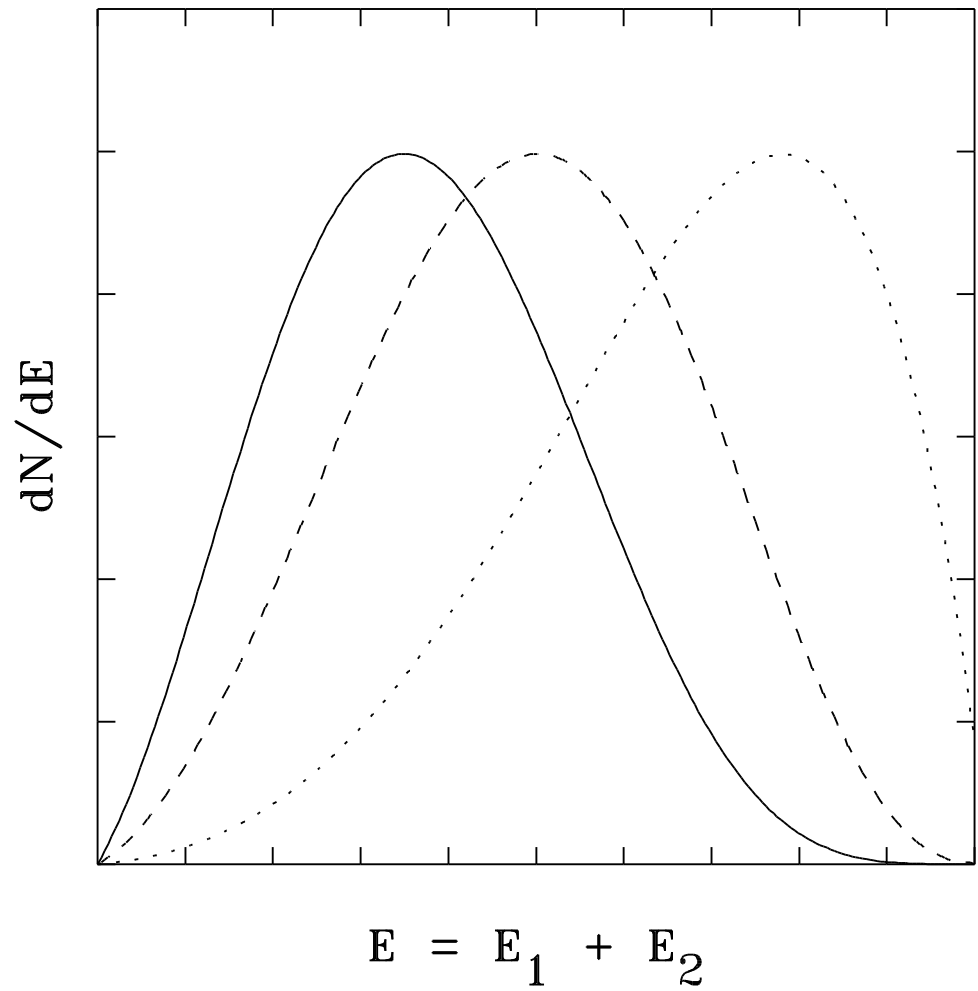
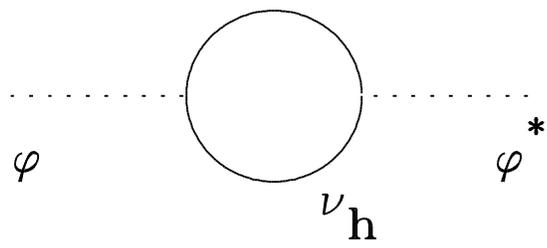
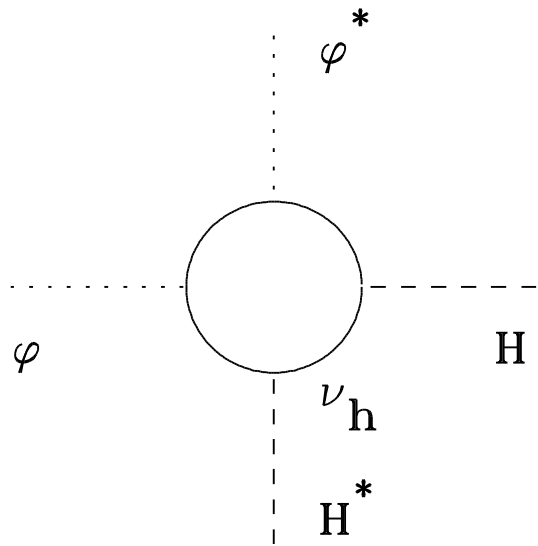
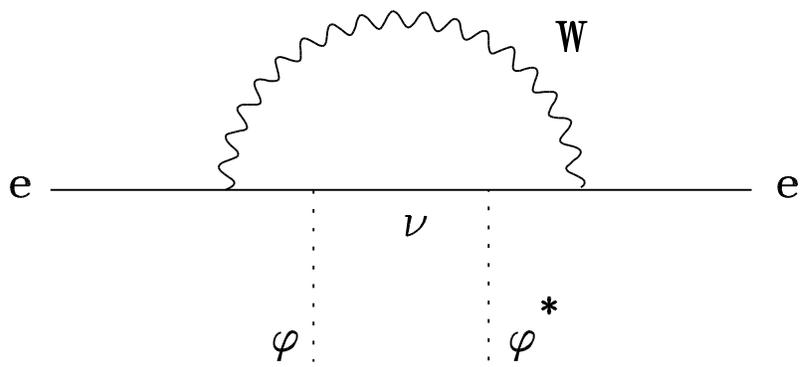
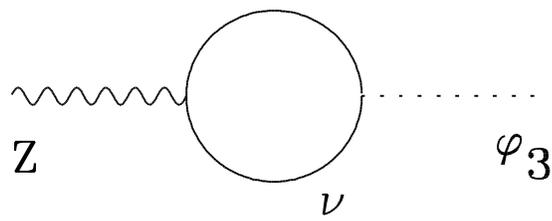
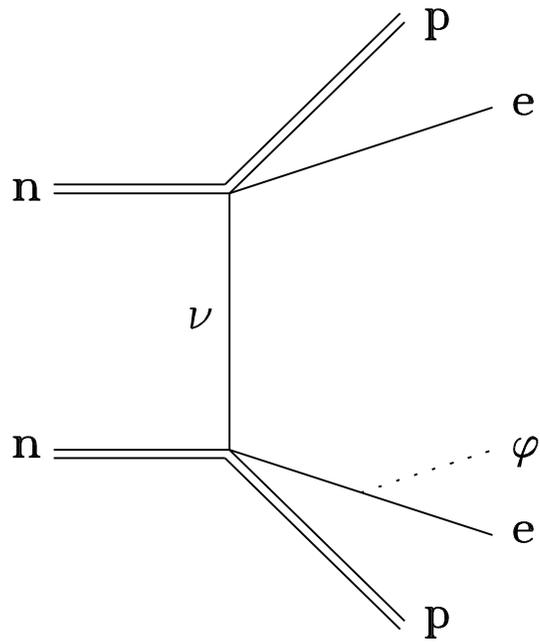
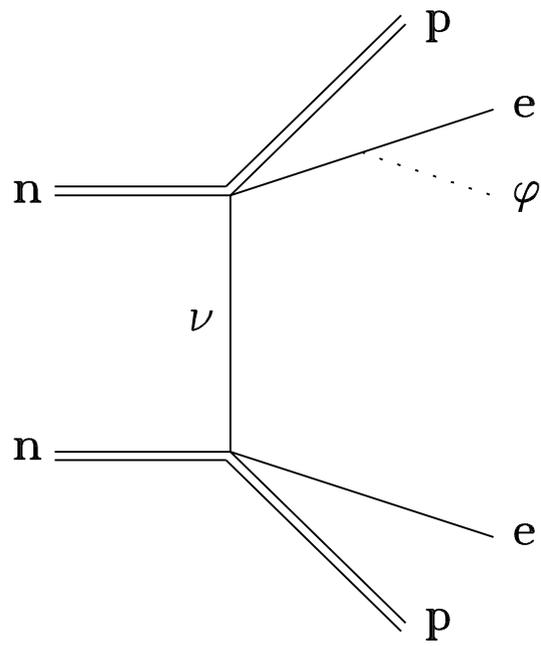
\bye